\documentclass[english,11pt,final,3p]{article}
\usepackage[T1]{fontenc}
\usepackage[latin9]{inputenc}
\usepackage{geometry}
\usepackage{xcolor}
\usepackage{float}
\usepackage{mathrsfs}
\usepackage{amsmath}
\usepackage{amssymb}
\usepackage{graphicx}
\usepackage[authoryear]{natbib}
\PassOptionsToPackage{normalem}{ulem}
\usepackage{ulem}

\makeatletter

\providecolor{lyxadded}{rgb}{0,0,1}
\providecolor{lyxdeleted}{rgb}{1,0,0}

\DeclareRobustCommand{\lyxsout}[1]{\ifx\\#1\else\sout{#1}\fi}

\newcommand{\lyxaddress}[1]{
	\par {\raggedright #1
	\vspace{1.4em}
	\noindent\par}
}

\usepackage{graphicx}
\usepackage{subfigure}
\usepackage{amsmath, amsthm, amssymb}

\makeatother

\usepackage{babel}
\begin{document}
\title{Notes on Propagation of 3D Buoyant Fluid-Driven Cracks}
\author{Dmitry I. Garagash and Leonid N. Germanovich}
\maketitle

\lyxaddress{Department of Civil and Resource Engineering, Dalhousie University,
1360 Barrington Street, Halifax, Nova Scotia B3H 4R2, Canada\\
Environmental Engineering and Earth Science Department, Clemson University,
Clemson, SC, USA}
\begin{abstract}
Magma-driven fractures is the main mechanism for magma emplacement
in the crust. A fundamental question is how the released/injected
fluid controls the propagation dynamics and fracture geometry (depth
and breadth) in three dimensions. Analog experiments in gelatin \citep[e.g. ][]{HeimpelOlson94,TaisneTait09}
show that fracture breadth (the short horizontal dimension) remains
nearly stationary when the process in the fracture \textquotedblleft head\textquotedblright{}
(where breadth is controlled) is dominated by solid toughness, whereas
viscous fluid dissipation is dominant in the fracture tail. We model
propagation of the resulting gravity-driven (buoyant), finger-like
fracture of stationary breadth with slowly varying opening along the
crack length. The elastic response to fluid loading in a horizontal
cross-section is local and can be treated similar to the classical
Perkins-Kern-Nordgren (PKN) model of hydraulic fracturing \citep{PeKe61,Nord72}.
The propagation condition for a finger-like crack is based on balancing
the global energy release rate due to a unit crack extension with
the rock fracture toughness. It allows us to relate the net fluid
pressure at the tip to the fracture breadth and rock toughness. Unlike
laterally propagating PKN fracture, where breadth is known a priori,
the final breadth of a finger-like vertically ascending fracture is
a result of processes in the fracture head. Because the head is much
more open than the tail, viscous pressure drop in the head can be
neglected leading to a 3D analog of Weertman\textquoteright s \citeyearpar{Weertman71}
hydrostatic pulse. This requires relaxing the local elasticity assumption
of the PKN model in the fracture head. As a result, we resolve the
breadth, and then match the viscosity-dominated tail with the 3-D,
toughness-dominated head to obtain a complete closed-form solution.
We then analyze the buoyancy-driven fracture propagation in conditions
of either continuous injection or finite volume release for sets of
parameters representative of low viscosity magma diking.
\end{abstract}

\section{Introduction}

Buoyancy magma-driven fractures (dikes) serve as the main mechanism
for magma ascension from deep sources and emplacement into the crust.
Modeling of such fractures is a long standing topic, see for example
reviews of \citep{Rubi95,ListerKerr91,Rivalta15_diketoughness}. Most
modeling efforts have focused on 2D, assuming plane-strain infinite-breadth
fracture, show that dikes establish a structure consisting of a buoyant,
nearly hydrostatic `head' with relatively enlarged opening connecting
to a thinner `tail' \citep{RoperLister07,SpenceTurcotte90,Lister90}.
For low enough magma injection rate at the source or a seized injection,
i.e. the fixed magma volume release, the solution in the dike head
is dominated by the solid toughness (and viscous pressure drop there
is negligible), while the solution in the tail is dominated by the
losses in the viscous magma flow. And it is the dynamics of magma-flow
in the tail that governs the ascent of the dike.

These modeling observations in 2D seem to be supported by dynamics
of laboratory dikes in 3D experiments of fluid-driven fracture of
gelatin \citep{Taka90,HeimpelOlson94,TaisneTait09}. Yet, given an
infinite breadth assumption, the 2D models remain hardly a practical
predictive or nature-interpretive tool, as they do not allow to link
dike propagation to emplaced magma-volumes. Furthermore, buoyant dikes
likely have a small breadth-to-length ratio once propagated out of
their source region, thus invalidating the plane-strain assumption
of 2D models.

\begin{figure}[tb]
\begin{centering}
(a)\includegraphics[scale=0.15]{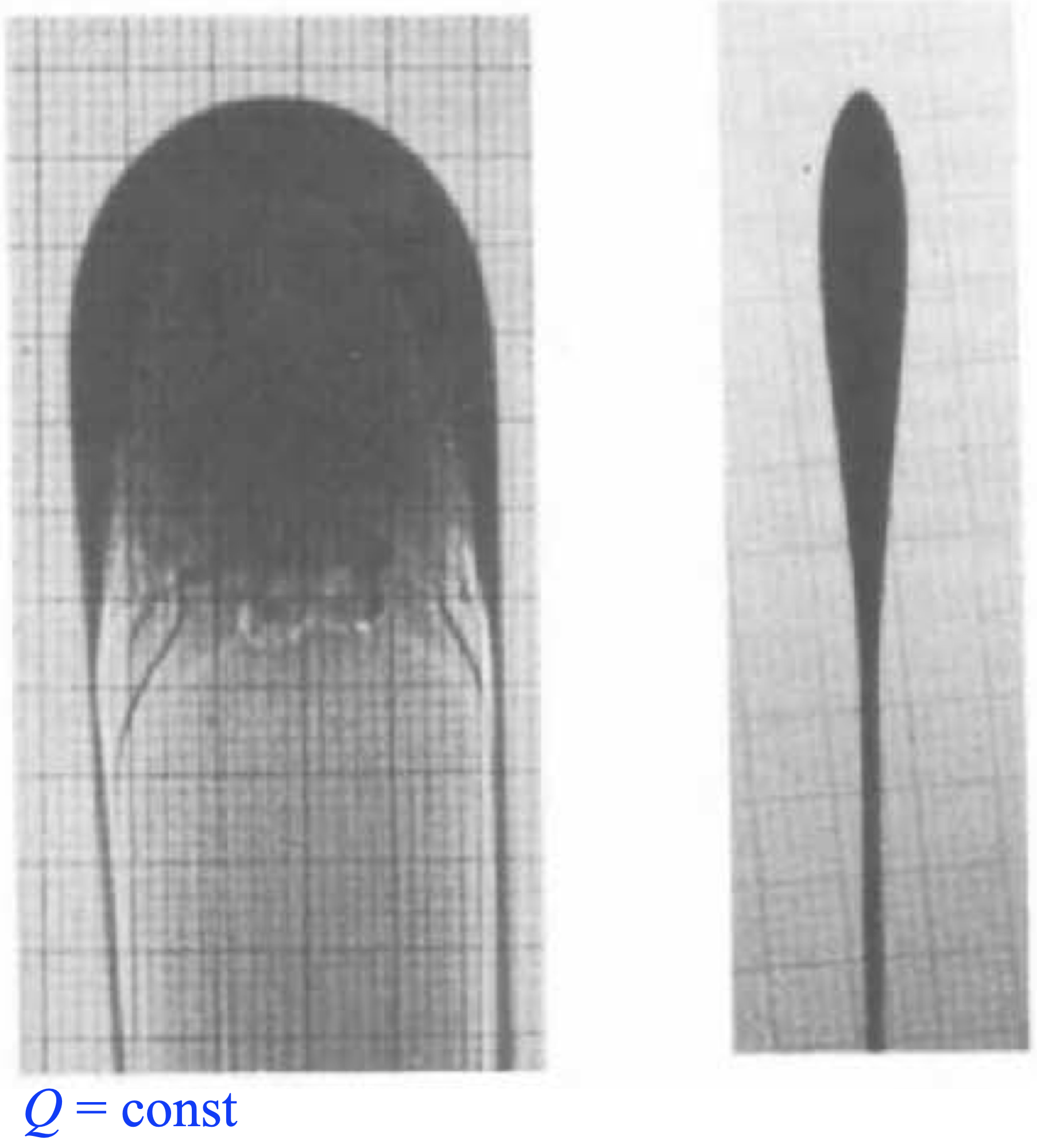}\hspace{0.5cm}(b)\includegraphics[scale=0.15]{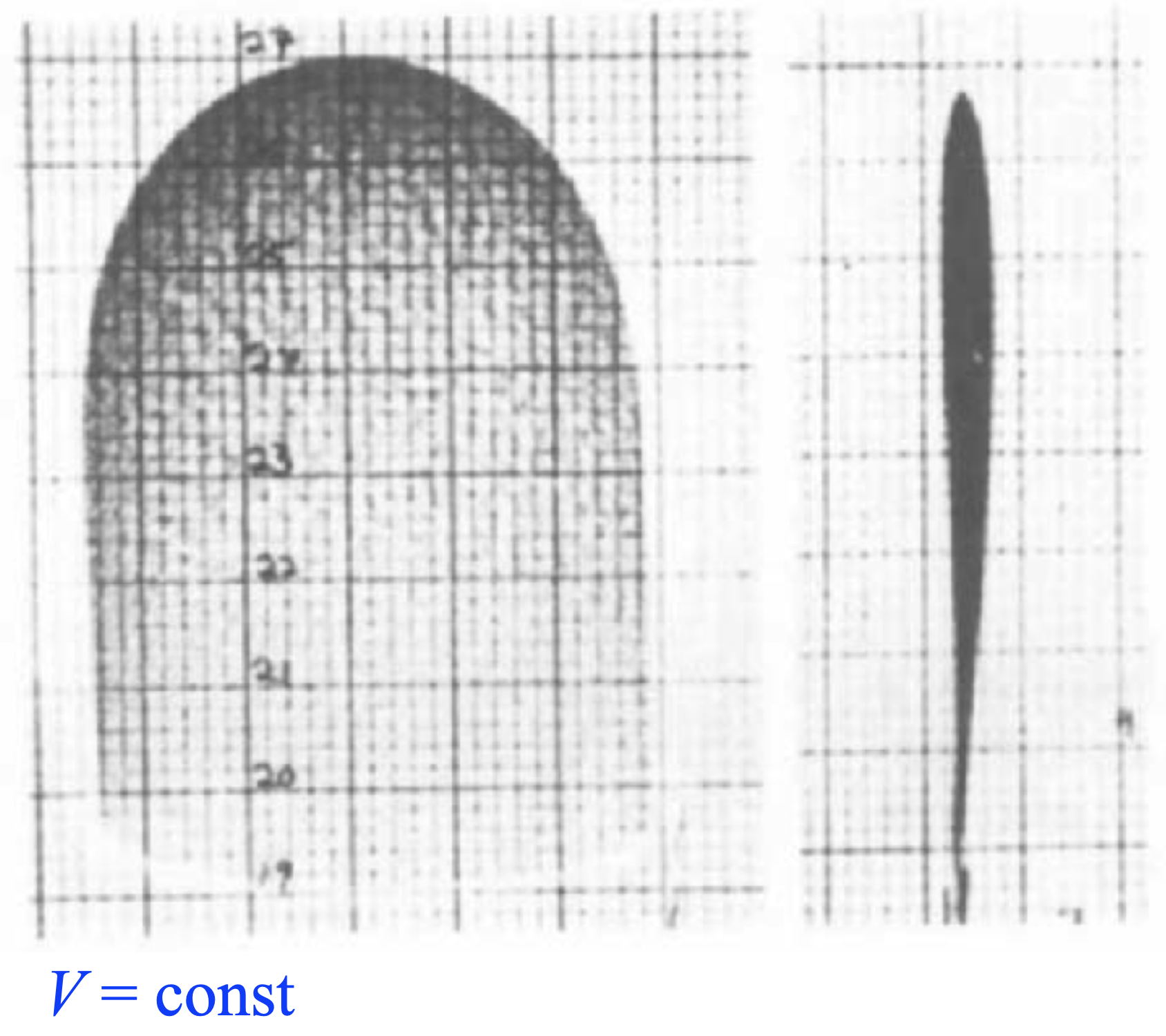}\vspace{0.5cm}
\par\end{centering}
\begin{centering}
(c)\includegraphics[scale=0.185]{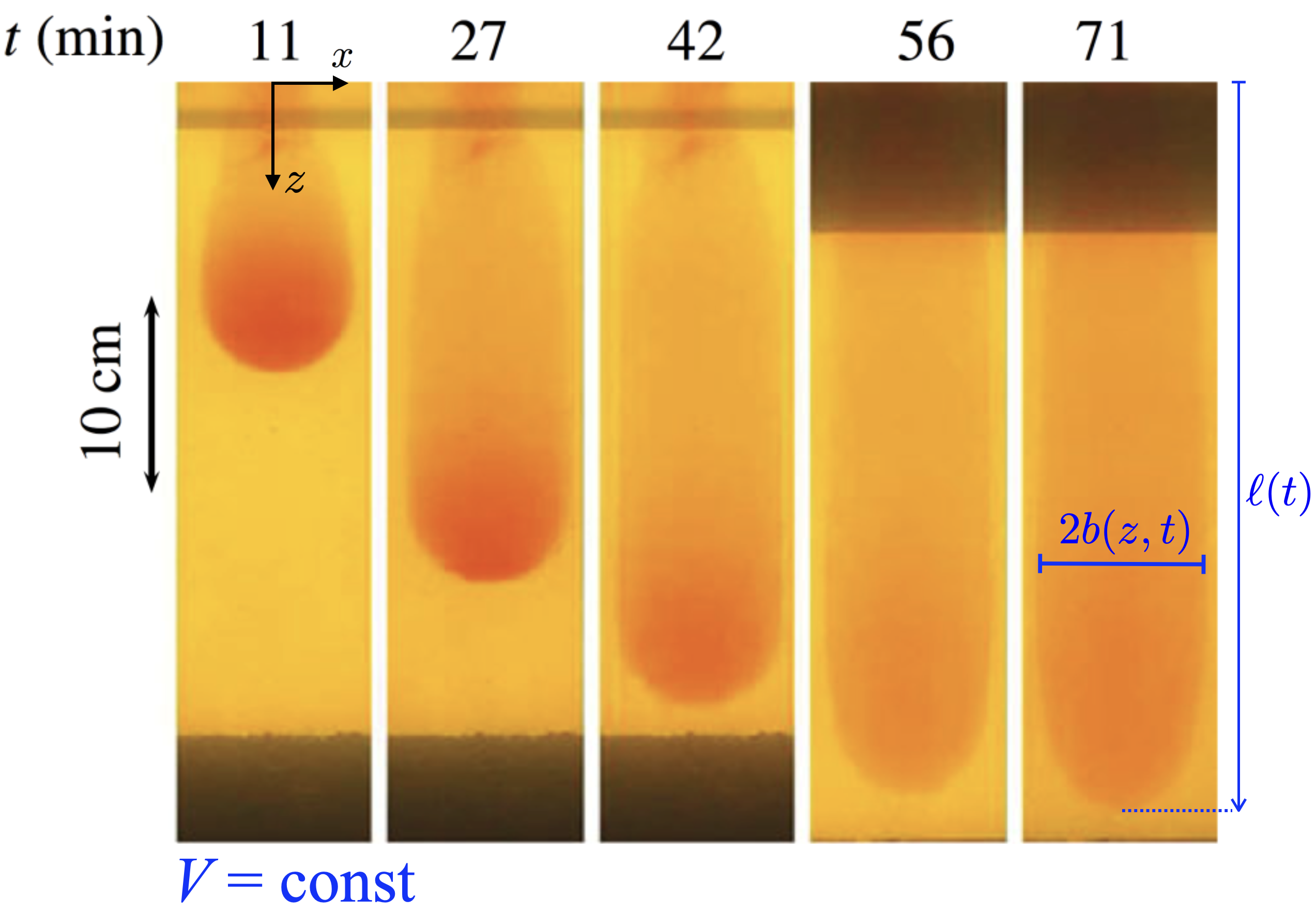}
\par\end{centering}
\caption{Examples of experimental dikes in gelatin with nearly constant breadth.
Dikes driven by constant rate of injection $Q$ of water solution
of cadmium chloride, \textbf{(a)}, and constant released volume $V$
of air, \textbf{(b)}, after \citet{HeimpelOlson94}. Snapshots of
breadth vs. elevation (left) and opening vs. elevation (right) are
shown. \textbf{(c)} `Reverse' dike driven by constant released volume
of sugar solution after \citet{taisne2011-dike-breadth,TaisneTait09}.
Snapshots of breadth vs. elevation at different propagation instances
are shown. \label{fig:exp}}
\end{figure}

In this paper we present a mechanical model of a 3D buoyant hydraulic
fracture with spontaneously developing breadth, as was first reported
by \citet{garagash2014gravity,Germanovich14AGU}. The model takes
a cue from observation of experimental dikes \citep{Taka90,HeimpelOlson94,TaisneTait09}
that the fracture breadth is generated by the fracturing process in
the head of the dike, and remain nearly stationary in the tail region
behind the head (Fig \ref{fig:exp}). 

We start in Section 2 of the paper with formulating a full 3D model
for a buoyant fluid-driven crack, followed in Section 3 with the reduction
of this model for the case of a small breadth-to-length ratio, similar
to the so-call PKN model for non-buoyant lateral propagation of finger-like
(or blade) fluid-driven cracks. Here we provide a general evolution
of this PKN-like model for a finger-dike, as well as, develop the
asymptotic, 'large-toughness' head-tail structure, with the explicit
asymptotic solutions for the toughness-dominated, hydrostatic head
and viscosity-dominated tail. Stationary breadth of the finger-like
dike cannot be determined within the PKN-like model, and have to be
prescribed there. Section 4 therefore develops the explicit 3D solution
for the hydrostatic dike head (Weertman's pulse in 3D), including
that for the dike breadth, which can be used to inform the PKN-like
model of Section 3. Comparison of the dike-head solution to the experiments
are also discussed. Section 5 follows with examples of the finger-dike
propagation solutions in the PKN-like approximation with breadth informed
by the dike-head solution of Section 4. Main results of the paper
are summarized in Section 6. 

\section{Formulation}

\begin{figure}[tb]
\begin{centering}
\includegraphics[scale=0.25]{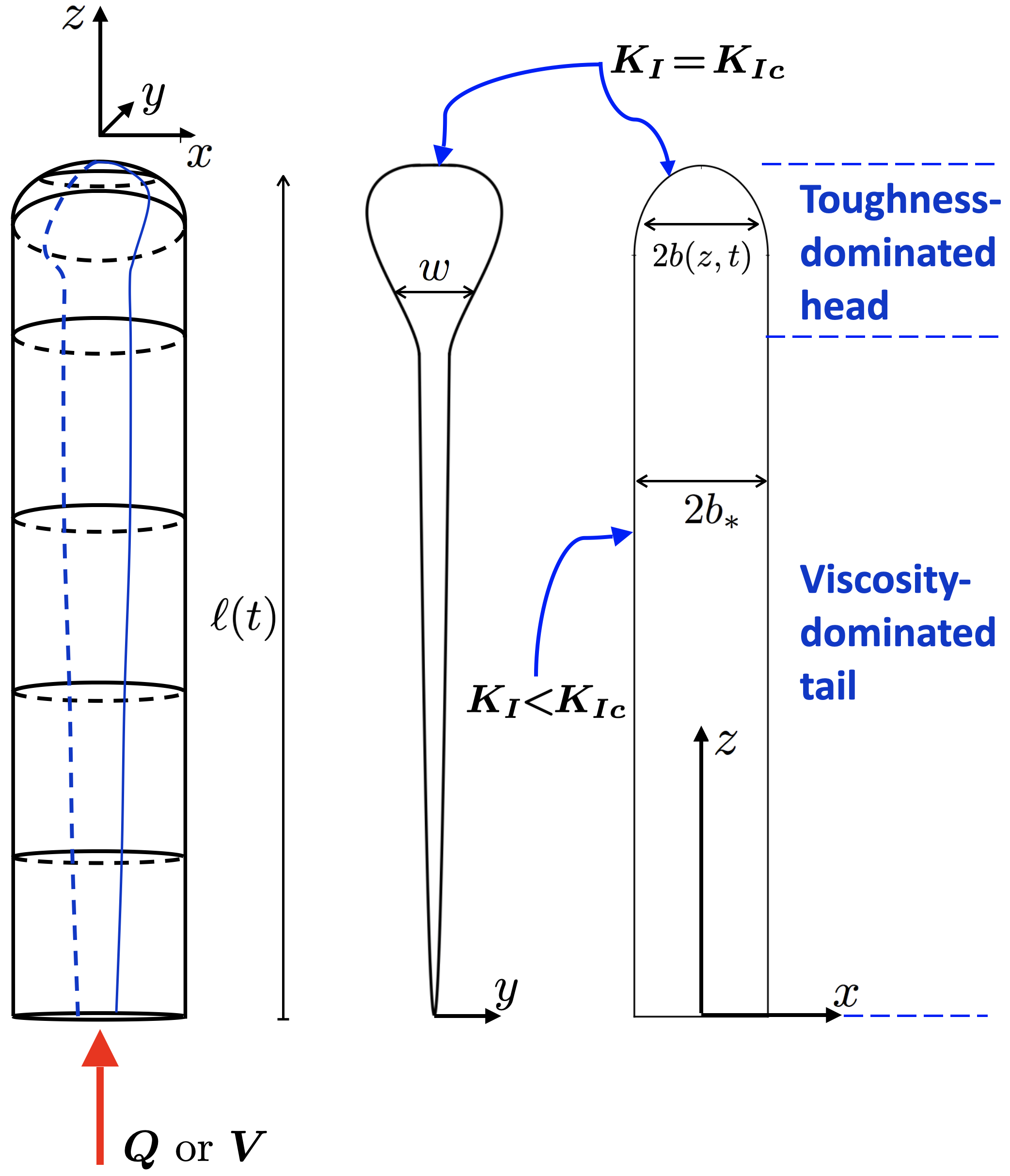}
\par\end{centering}
\caption{\textbf{(a)} Model of 3D dike propagation with small aspect ratio
(breadth-to-length) in the buoyancy direction $z$ due to injection
of fluid at the inlet $z=0$. \textbf{(b)} Characteristic variation
of the maximum dike opening $w(x=0,z)$ with elevation $z$ (view
in the direction of the dike breadth, along $x$-axis). \textbf{(c)}
Characteristic variation of the dike half-breadth $b(z)$ with elevation
$z$ (view in the direction of the dike opening, along $y$-axis).
Along its length, dike is schematically separated into the toughness-dominated
`head', where the rock is fractured by advancing dike to widen the
half-breadth $b(z)$ from at the dike tip $z=\ell(t)$ to the maximum
value $b_{*}$, and the viscosity-dominated `tale', where the breadth
remains saturated. \label{fig:model}}
\end{figure}

We consider a crack propagating in the fluid ``buoyancy direction'',
$(\rho_{f}-\rho_{s})\mathbf{g}$ ($\mathbf{g=}$ the gravity vector),
aligned here with the $z$-axis, from a source at the origin of the
coordinate frame (Fig \ref{fig:model}). The crack plane is perpendicular
to the direction of the minimum in-situ stress ($y$-axis), which
is assumed to vary along along the lithostatic gradient, $\sigma_{yy}^{\infty}=\pm\rho_{s}gz+\sigma_{o}$,
where '$\pm$' corresponds to the sense of buoyancy, $\rho_{f}-\rho_{s}\gtrless0$,
and $\sigma_{o}$ is the reference stress level corresponding to the
value at $z=0$. (Stress is positive in compression). The fracture
surface $\mathcal{S}$ is specified by $|x|\le b(z,t)$ with $0\leq z\le\ell(t)$,
where $2b(z,t)$ is the local breadth of the crack and $\ell(t)$
is its length. 

It proves convenient to make use of the following effective material
parameters 
\begin{equation}
\bar{K}=\sqrt{\frac{2}{\pi}}K_{Ic}\qquad\bar{E}=\frac{1}{\pi}\frac{E}{1-\nu^{2}}\qquad\bar{\mu}=\pi^{2}\mu\label{K'}
\end{equation}
related by a numerical factor to the solid toughness $K_{Ic}$, the
solid plane-strain elastic modulus $E/(1-\nu^{2})$, and the dynamic
fluid viscosity $\mu$.

The crack opening $w=w(x,z,t)$ (= displacement discontinuity in the
$y$-direction) is related to the net fluid pressure in the crack,
$p=p_{f}(x,z,t)-\sigma_{yy}^{\infty}(z)$, by a non-local relation
of elasticity \citep[e.g. ][]{HiKe96}

\begin{equation}
p(x,z)=-\frac{\bar{E}}{8}\int_{\mathcal{S}}\frac{w(x',z')}{[(x-x')^{2}+(z-z')^{2}]^{3/2}}dx'dz'\label{elast}
\end{equation}
where $\bar{E}$ is an elastic modulus, (\ref{K'}), and time $t$
was suppressed from the list of arguments for brevity. The integral
in (\ref{elast}) is hypersingular and is understood in the Hadamard
sense.

The flow of incompressible Newtonian fluid in the crack is governed
by the local continuity, 
\begin{equation}
\frac{\partial w}{\partial t}+\mathrm{div}(w\mathbf{v})=0,\label{cont}
\end{equation}
and by the Poiseuille law for the flow velocity $\mathbf{v}=\{v_{x},0,v_{z}\}$,
\begin{equation}
\mathbf{v}=-\frac{w^{2}}{12\mu}\nabla\left(p-\Delta\rho gz\right),\label{v}
\end{equation}
where $\Delta\rho=|\rho_{f}-\rho_{s}|$. We further assume slowly
varying (if at all) breadth of a buoyancy-driven fracture, and, thus,
negligible lateral components of fluid velocity and pressure gradient,
\begin{equation}
v_{x}=\partial p/\partial x=0.\label{vx}
\end{equation}
This assumption may possibly breakdown at early stages of the fracture
when its breadth and length are comparable. 

We further assume that the fluid wets the entire crack (i.e. the lag
between the fluid and the fracture edge is negligible \citep[e.g. ][]{Rubi93,GaragashDetournay00}
and the fluid velocity there is identical to the edge velocity)
\begin{equation}
v=\pm\frac{\partial b/\partial t}{\sqrt{1+(db/dz)^{2}}}\quad\text{at}\quad x=\pm b\label{edge}
\end{equation}
Under this assumption and assuming that the leak-off into the rock
is negligible, the fluid source condition can be equivalently formulated
as the statement of global fluid continuity, 
\begin{equation}
V(t)=\int_{\mathcal{S}}w(x,z,t)dxdz\label{global}
\end{equation}
where $V(t)$ is the cumulative volume of the injected fluid.

Adopting linear elastic fracture mechanics, the fracture propagation
in mobile equilibrium requires that the local opening-mode stress
intensity factor $K_{I}(z,t)$ along the fracture edge $|x|=b(z,t)$
is equal to or less than the solid toughness $K_{Ic}$ along the propagating
and stationary parts of the front, respectively,
\begin{equation}
K_{I}(z,t)=K_{Ic}\quad(\partial b/\partial t>0),\qquad K_{I}(z,t)<K_{Ic}\quad(\partial b/\partial t=0)\label{KI}
\end{equation}
This propagation condition can also be conveniently rephrased in terms
of the asymptotic behavior of the crack opening near (the propagating
part of) the crack edge as \citep{Irvin57} 
\begin{equation}
w_{\perp}(r)=\frac{4}{\pi}\frac{\bar{K}}{\bar{E}}\sqrt{r}\qquad(r\rightarrow0)\label{wn}
\end{equation}
where $\bar{K}$ is a toughness parameter, (\ref{K'}), and $w_{\perp}(r)$
in the crack opening in a cross-section normal to the front ($r$
is the distance from the front within the cross-section). Along the
non-propagating part of the crack edge, $w_{\perp}(r)<\frac{4}{\pi}\frac{\bar{K}}{\bar{E}}\sqrt{r}$. 

\section{Propagation of Finger Crack with Stationary Breadth (PKN-Approximation)\label{sec:PKN}}

\subsection{Preliminaries }

It has been observed in laboratory experiments in gelatin \citep{Taka90,HeimpelOlson94,TaisneTait09}
that the fracture breadth is nearly stationary when the fracturing
process near the fracture head ($z=\ell(t)$), that ``generates''
the final breadth of the crack $b_{*}$, is dominated by the solid
toughness, as opposed to the dissipation in the viscous fluid flow
in the crack. This is symptomatic of buoyancy-driven cracks which
tend to have maximum opening in the head region, which, in turn, minimizes
the viscous losses there. 
\begin{figure}[tb]
\begin{centering}
\includegraphics[scale=0.4]{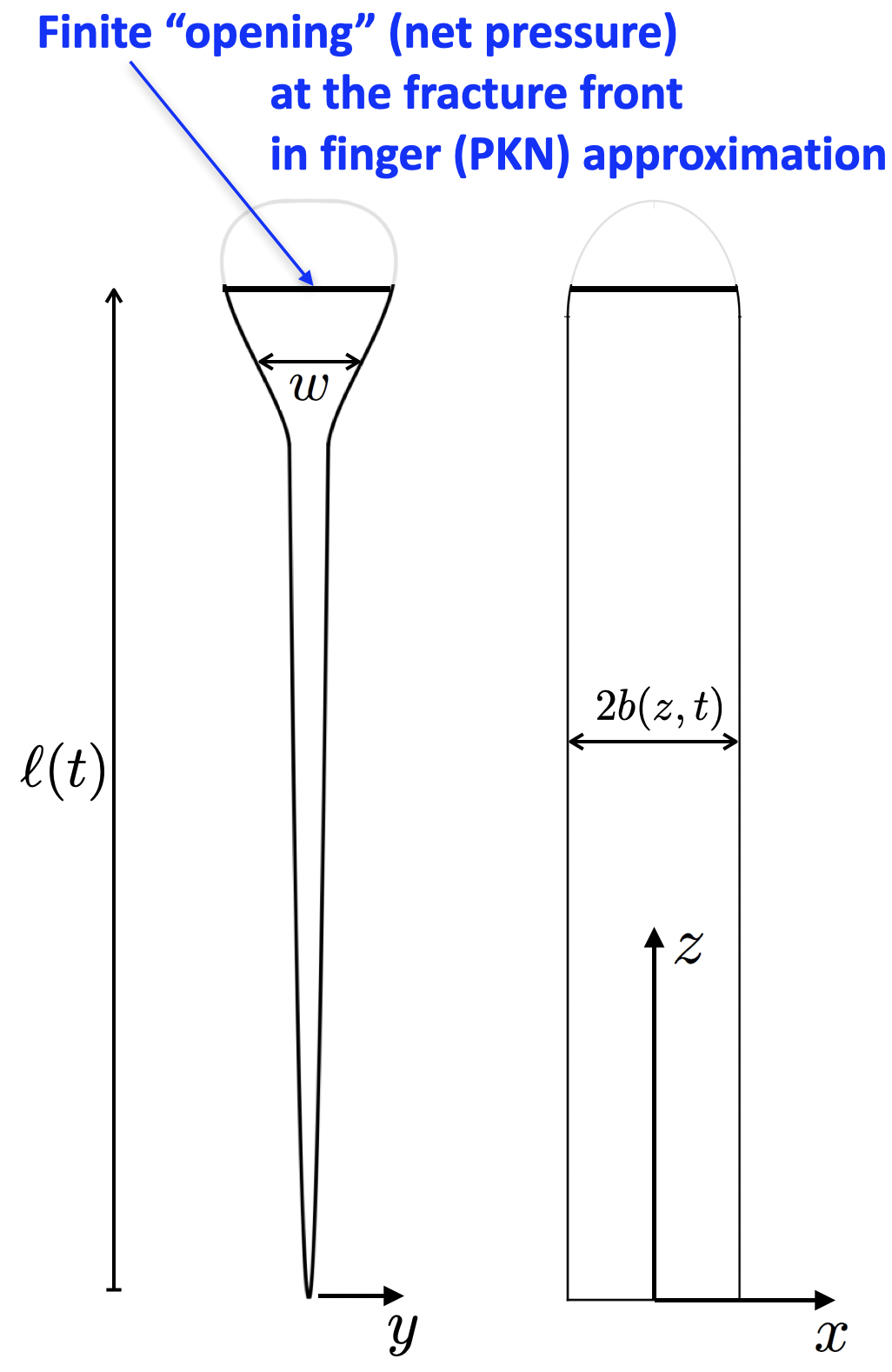}
\par\end{centering}
\caption{PKN approximation, where small aspect ratio 3D fracture is modeled
by local (in the opening-breadth cross-section) elasticity. Energy-based
propagation condition corresponds to finite opening at the PKN crack
tip prescribed by the material fracture toughness. \label{fig:PKN}}
\end{figure}

We turn to study such buoyancy-driven cracks with a small aspect (breadth-to-length)
ratio,
\begin{equation}
b_{*}\ll\ell(t)\label{PKN}
\end{equation}
The determination of the actual value of the stationary breadth $b_{*}(z)$
will require explicit consideration of the structure of the ``head
region'', which is delayed until the next section. We only note at
this point that the breadth is expected to be invariant with depth
(fracture height) if the rock properties are depth-independent. Conversely,
if the fracture toughness and/or the rock modulus and/or rock density
(and thus density mismatch between the fluid and the rock) are depth
dependent (which is conceivable if kilometers high dikes/fracture
are considered), then the breadth is expected to vary with depth.
Thus, in what follows, we allow for the rock modulus and the rock
toughness are to be some predetermined functions of depth.

For a finger-like crack, $b_{*}\ll\ell$, and ``slowly'' varying
opening along the crack length, $\partial w/\partial z\sim w/\ell$,
the convolution kernel in (\ref{elast}) can be approximated as 
\[
\frac{1}{[(x-x')^{2}+(z-z')^{2}]^{3/2}}\approx\frac{2}{(x-x')^{2}}\delta_{Dirac}(z-z')
\]
which then allows to reduce (\ref{elast}) to its plane-strain equivalent
\citep[e.g.][]{BiEs68}
\begin{equation}
p(x,z)=-\frac{\bar{E}(z)}{4}\int_{-b_{*}(z)}^{b_{*}(z)}\frac{w(x',z)}{(x-x')^{2}}dx'\label{elast-1}
\end{equation}
Once again, this approximation breaks down in the vicinity of the
crack ``tips'' (i.e., near $z=0$ and $z=\ell$). 

Since the pressure is approximately equilibrated along the breadth,
(\ref{vx}), $p(x,z)=p(z)$, eq. (\ref{elast-1}) yields classical
elliptical crack opening that can be expressed as
\begin{equation}
w(x,z)=\frac{4\overline{w}(z)}{\pi}\sqrt{1-\frac{x^{2}}{b_{*}^{2}(z)}},\label{w-1}
\end{equation}
where $\overline{w}(z)$ is the breadth-averaged opening
\begin{equation}
\overline{w}(z)=b_{*}(z)\frac{p(z)}{\bar{E}(z)},\label{wbar}
\end{equation}
and time $t$ was omitted from the list of arguments in $w$, $\overline{w}$,
and $p$. 

Integrating the fluid flow equations along the crack breadth (and
using the zero lag condition (\ref{edge})) yields
\begin{equation}
\frac{\partial b_{*}\overline{w}}{\partial t}+\frac{\partial b_{*}\overline{q}}{\partial z}=0,\label{contbar}
\end{equation}
where $\overline{q}=\overline{wv}$ is the cross-section averaged
flow rate
\begin{equation}
\overline{q}=-\frac{\overline{w}^{3}}{\bar{\mu}}\left(\frac{\partial p}{\partial z}-\Delta\rho g\right)\label{qbar}
\end{equation}
where $\bar{\mu}=\pi^{2}\mu$ is a dynamic viscosity parameter, as
defined earlier in (\ref{K'}). 

The zero fluid lag condition applied at the ``tip'' yields 
\begin{equation}
\lim_{z\rightarrow\ell}(\overline{q}/\bar{w})=d\ell/dt\label{tip}
\end{equation}
while the global fluid volume balance reduces to:
\begin{equation}
V=\int_{0}^{\ell}2b_{*}\overline{w}dz\label{globalbar}
\end{equation}

The closure of the approximate model (\ref{w-1}-\ref{globalbar}),
which is in so far equivalent to the classical PKN model of a hydraulic
fracture with restricted breadth \citep{PeKe61,Nord72,Kemp90}, requires
specifying an additional constraint or boundary condition. Naturally,
the latter has to do with the crack propagation condition. Unfortunately,
since the solution in the head of the fracture is not resolved by
the current approximation, the propagation condition in the form (\ref{KI})
can not be applied directly. Instead, the ``net'' propagation criterion
for the fracture head can be established from estimating the global
energy release rate in extending the finger crack and equating it
to the fracture energy $K_{Ic}^{2}/E'$ \citep{SarvaraminiGaragash15,Garagash22}.
This ``net'' propagation constraint is:
\begin{equation}
p=\frac{\bar{K}}{\sqrt{b_{*}}}\quad\text{at}\quad z=\ell\label{prop}
\end{equation}
We emphasize again that the approximate model of a finger crack includes
the crack breadth $b(z)$. Therefore, when $b(z)$ is known a priori,
such as in the case of a hydraulic fracture propagation in a layer
sandwiched between tougher/stiffer layers and/or layers subjected
to higher compressive stress, this model is complete \citep{SarvaraminiGaragash15,Garagash22}.
Otherwise, as is true for a buoyant crack in a homogeneous rock, $b_{*}(z)$
is not known a priori, and is a part of the solution in the fracture
head (Section \ref{sec:head}).

\subsection{Scaling (depth-independent rock properties and fracture breadth)}

\textbf{}

Let us nondimensionalize the field variables with respect to the scales
\begin{align}
x_{*} & =z_{*}=L_{*}\quad t_{*}=\frac{L_{*}}{v_{*}}\quad w_{*}=\frac{\bar{K}\sqrt{L_{*}}}{\bar{E}}\quad p_{*}=\frac{\bar{K}}{\sqrt{L_{*}}}\nonumber \\
 & v_{*}=\left(\frac{\bar{K}}{\bar{E}}\right)^{2}\frac{p_{*}}{\bar{\mu}}\quad V_{*}=w_{*}L_{*}^{2}\quad Q_{*}=\frac{V_{*}}{t_{*}}\label{norm}
\end{align}
which are predicated on a single lengthscale $L_{*}$, chosen here
as the buoyancy length \citep{ListerKerr91}
\begin{equation}
L_{*}=\left(\frac{\bar{K}}{\Delta\rho g}\right)^{2/3}\label{Lb}
\end{equation}
Expanded form of the scales (\ref{norm}) with (\ref{Lb}) is:
\begin{align}
x_{*} & =z_{*}=\left(\frac{\bar{K}}{\Delta\rho g}\right)^{2/3}\quad t_{*}=\frac{\bar{\mu}\bar{E}^{2}}{\Delta\rho g\bar{K}^{2}}\quad w_{*}=\frac{\bar{K}^{4/3}}{\bar{E}\,(\Delta\rho g)^{1/3}}\quad p_{*}=\left(\Delta\rho g\bar{K}^{2}\right)^{1/3}\nonumber \\
 & v_{*}=\frac{\bar{K}^{8/3}(\Delta\rho g)^{1/3}}{\bar{\mu}\bar{E}^{2}}\quad V_{*}=\frac{\bar{K}^{8/3}}{\bar{E}\,(\Delta\rho g)^{5/3}}\quad Q_{*}=\frac{\bar{K}^{14/3}}{\bar{\mu}\bar{E}^{3}(\Delta\rho g)^{2/3}}
\end{align}

Elasticity, fluid continuity and the Poiseuille's equations can be
rewritten in units of (\ref{norm}), or, conversely, in the normalized
form as
\begin{equation}
\overline{w}=b_{*}\,p\qquad\frac{\partial\overline{w}}{\partial t}=-\frac{\partial\overline{q}}{\partial z}\qquad\overline{q}=-\overline{w}^{3}\left(\frac{\partial p}{\partial z}-1\right)\label{PDE}
\end{equation}

Plugging the first and the third into the second in (\ref{PDE}) we
have
\begin{equation}
\frac{\partial p}{\partial t}=b_{*}^{2}\frac{\partial}{\partial z}\left[p^{3}\left(\frac{\partial p}{\partial z}-1\right)\right]\label{PDE_p}
\end{equation}
This non-linear PDE is subjected to the following global and boundary
conditions:
\begin{equation}
V=2b_{*}^{2}\int_{0}^{\ell}pdz\quad\text{or}\quad2b_{*}\bar{q}_{|z=0}=\frac{dV}{dt}\label{V}
\end{equation}
\begin{equation}
(\overline{q}/\bar{w})_{|z=\ell}=\frac{d\ell}{dt}\quad\text{and}\quad p_{|z=\ell}=\frac{1}{\sqrt{b_{*}}}\label{tip-1}
\end{equation}
The normalized solution $p(z,t)$ and $\ell(t)$ of (\ref{PDE_p}-\ref{tip-1})
depends on the normalized injected fluid volume $V(t)$, and normalized
fracture breadth $b_{*}$. 

The formulation for the case when the rock properties and the fracture
breadth vary with depth is given in Appendix \ref{app:depth}.

\subsection{Normalized Solution using Scales (\ref{norm})}

Numerical method of lines is used to solve the non-linear PDE (\ref{PDE_p})
with boundary conditions (\ref{V}) and (\ref{tip-1}), as detailed
in Appendix \ref{App:PKN}. The numerical solution reveals at large
enough times an asymptotic structure comprised of a hydrostatic head
region, where viscous stresses are negligible compared to the elastic
ones and solution is stationary in the frame moving with the crack
tip, and a viscous tail region, where the elastic stress is negligible.
These two asymptotic head and tail solutions are discussed below.

\subsubsection{Outer Solution for the Buoyant (Hydrostatic) Head}

Hydrostatic net pressure

\begin{equation}
p(z,t)=z-\ell_{\mathrm{tail}}(t)\qquad(\ell_{\mathrm{tail}}<z<\ell)\label{phead}
\end{equation}
Applying the propagation condition at the tip ($p_{|z=\ell}=1/\sqrt{b_{*}}$)
, 
\begin{equation}
\ell_{\mathrm{head}}=\ell-\ell_{\mathrm{tail}}=1/\sqrt{b_{*}}\label{lhead}
\end{equation}
The volume of the head is 
\begin{equation}
V_{\mathrm{head}}=b_{*}^{2}\ell_{\mathrm{head}}^{2}=b_{*}\label{Vhead}
\end{equation}

In \emph{dimensional} terms, i.e. restoring ``units'' (\ref{norm})
in the above two expressions, we have
\[
\ell_{\mathrm{head}}=\frac{1}{\sqrt{b_{*}/L_{*}}}L_{*},\qquad V_{\mathrm{head}}=\frac{b_{*}}{L_{*}}\,\frac{\bar{K}L_{*}^{5/2}}{\bar{E}}
\]

\subsubsection{Asymptotic Solution for the Viscous Tail}

This is the case analyzed for a plane-strain (infinite breadth) dyke
by \citet{SpenceSharp87} and \citet{Lister90} in the case of constant
fluid injection rate and by \citet{SpenceTurcotte90} in the case
of constant injected fluid volume. The solution for a finite breadth
crack considered here is different from these plane-strain solutions
by numerical prefactors only. In the viscous tail, $0<z<\ell_{\mathrm{tail}}(t)$,
the net pressure gradient is negligible compared to the buoyancy gradient
over most of the fracture length, i.e. $|\partial p/\partial z|\ll1$,
and a similarity solution can be obtained by usual methods in the
form $\ell_{\mathrm{tail}}\sim t^{(2\alpha+1)/3}$, $p=t^{(\alpha-1)/3}\Pi(z/\ell_{\mathrm{tail}}(t))$
for an arbitrary injection power-law $V\sim t^{\alpha}$ with $\alpha\ge0$.
The solutions for the two cases of most interest are given below.

Constant injection rate $dV/dt=Q$:
\begin{equation}
p=\left(\frac{Q}{2b_{*}^{4}}\right)^{1/3}\qquad\ell_{\mathrm{tail}}=b_{*}^{2}p^{2}\Delta t=\left(\frac{Q}{2b_{*}}\right)^{2/3}\Delta t\qquad\left(\Delta t=t-\frac{V_{\mathrm{head}}}{Q}\right)\label{ptail1}
\end{equation}

Constant volume release $V=\mathrm{const}$:
\begin{equation}
p=\left(\frac{z}{3b_{*}^{2}t}\right)^{1/2}\qquad\ell_{\mathrm{tail}}=\left(\frac{27}{16b_{*}^{2}}V_{\mathrm{tail}}^{2}t\right)^{1/3}\qquad\left(p_{\mathrm{neck}}=\left(\frac{V_{\mathrm{tail}}}{4b_{*}^{4}t}\right)^{1/3}\right)\label{ptail2}
\end{equation}
where $V_{\mathrm{tail}}=V-V_{\mathrm{head}}$ is the constant volume
of the tail, and $p_{\mathrm{neck}}=p_{|z=\ell_{\mathrm{tail}}}$
is the maximum net pressure in the tail, at the juncture with the
head (we refer to as the ``neck''). 

Couple comments about these solutions are in order. Since the volume
of the fracture head ($V_{\mathrm{head}}$) is stationary, the volume
of the tail ($V_{\mathrm{tail}}$) expands at the rate of fluid injection.
The two solutions satisfy the tail continuity equation $V_{\mathrm{tail}}=2b_{*}^{2}\int_{0}^{\ell_{\mathrm{tail}}}pdz$
and the velocity condition at the juncture between the expanding tail
and the stationary head, $\overline{q}=\bar{w}\,(d\ell_{\mathrm{tail}}/dt)$
at $z=\ell_{\mathrm{tail}}(t)$.

Furthermore, in order for the asymptotic dike structure to be realizes
(i.e., hydrostatic head ``attached to'' the viscous tail with \emph{stationary}
breadth), the pressure in the viscous tail should not exceed the fracturing
value at the tip, $p\le1/\sqrt{b_{*}}$ (or in dimensional terms,
$p\le\sqrt{2}K_{Ic}/\sqrt{\pi b_{*}}$). More precisely, to avoid
expansion of the tail breadth, pressure there should not exceed the
``breadth fracturing'' value for a uniformly pressurized plane-strain
cross-section through (along?) the breadth, $p\le1/\sqrt{2b_{*}}$
(or in dimensional terms, $p\le K_{Ic}/\sqrt{\pi b_{*}}$). This requires,
that, in the injection case, the normalized injection rate does not
exceed a critical value
\begin{equation}
Q<\frac{b_{*}^{5/2}}{\sqrt{2}}\qquad(Q=\mathrm{const})\label{Q_pkn}
\end{equation}
and in the shut-in case, the normalized time exceeds the critical
one (when the pressure at $z=\ell_{tail}$ falls below the breadth
fracturing value)
\begin{equation}
\frac{V_{tail}}{t}<b_{*}^{5/2}\sqrt{2}\qquad(V=\mathrm{const})\label{t_pkn}
\end{equation}
The dimensional form of these constraints obtained by multiplying
the right hand sides by the injection rate scale $V_{*}/t_{*}=(\bar{K}{}^{4}/\bar{E}{}^{3})(L_{*}/\bar{\mu})$. 

In \emph{dimensional} terms, the net-pressure is related to the average
opening by (\ref{wbar}) and the above normalized asymptotic solutions
translate to the dimensional expressions for the tail length and width
using corresponding scales (\ref{norm}).  In the interest of comparing
these solutions to their two-dimensional (infinite breadth dike) counterparts,
we express the former in terms of the equivalent 2-D rate and 2-D
volume (area), respectively, 
\[
Q_{2D}=Q/(2b_{*})\qquad V_{2D}=V_{\mathrm{tail}}/(2b_{*})
\]
as
\[
\overline{w}=\left(\frac{\bar{\mu}Q_{2D}}{\Delta\rho g}\right)^{1/3}\qquad\ell_{tail}=\frac{Q_{2D}t}{\overline{w}}\qquad(Q=\mathrm{const})
\]
\[
\overline{w}=\left(\frac{\bar{\mu}}{3\Delta\rho g}\frac{z}{t}\right)^{1/2}\qquad\ell_{tail}=\left(\frac{27V_{2D}^{2}\Delta\rho gt}{4\bar{\mu}}\right)^{1/3}\qquad(V=\mathrm{const})
\]
These 3D dikes expressions are equivalent to the corresponding expressions
for plane-strain dikes \citep[e.g., equations (2.4) and (6.7) of][respectively]{RoperLister07}
upon replacing the viscosity parameter $\bar{\mu}=\pi^{2}\mu$ by
$12\mu$. This, for example, translates to a factor $(12/\pi^{2})^{1/3}\approx1.067$
difference between the values of the length of the viscous tail of
a finite- and an infinite- breadth dikes. 

\subsection{Asymptotic solution with partial head}

The asymptotic solution with the full hydrostatic head attached to
the viscous tail is realized when the neck pressure (opening), i.e.,
as previously defined maximum pressure in the tail occurring at the
juncture of the tail and the head, is negligible compared to the prevailing
pressure in the head, i.e. $p_{\mathrm{neck}}\ll1$. This asymptotic
framework can be approximately extended to the case when $p_{\mathrm{neck}}$
is not vanishingly small by ``attaching'' the partial head (``circumcised''
at the value of pressure equal to $p_{\mathrm{neck}}$) to the viscous
tail, such that the partial head size and volume are
\[
\ell_{\mathrm{head}}=\frac{1}{\sqrt{b_{*}}}-p_{\mathrm{neck}}\qquad V_{\mathrm{head}}=b_{*}-b_{*}^{2}p_{\mathrm{neck}}^{2}
\]
where $p_{\mathrm{neck}}$ is stationary for a constant injection
rate, (\ref{ptail1}), and decreasing with time for a constant volume
release, (\ref{ptail2}). In the latter case, since $V_{\mathrm{head}}$
evolves with time, so does $V_{\mathrm{tail}}$, which now satisfies
an implicit equation
\[
V-V_{\mathrm{tail}}=b_{*}-b_{*}^{2}\left(\frac{V_{\mathrm{tail}}}{4b_{*}^{4}t}\right)^{2/3}\qquad(V=\mathrm{const})
\]
Corresponding solution with non-stationary volume in the tail is no
longer strictly self-similar, yet, we hypothesize that the constant
$V_{tail}$ solution formally used with a slowly varying $V_{tail}$
may still yield an adequate approximation...

\section{The Full Solution for the Buoyant (Hydrostatic) Head\label{sec:head}}

\subsection{Full Numerical Solution}

\begin{figure}[tb]
\includegraphics[scale=0.35]{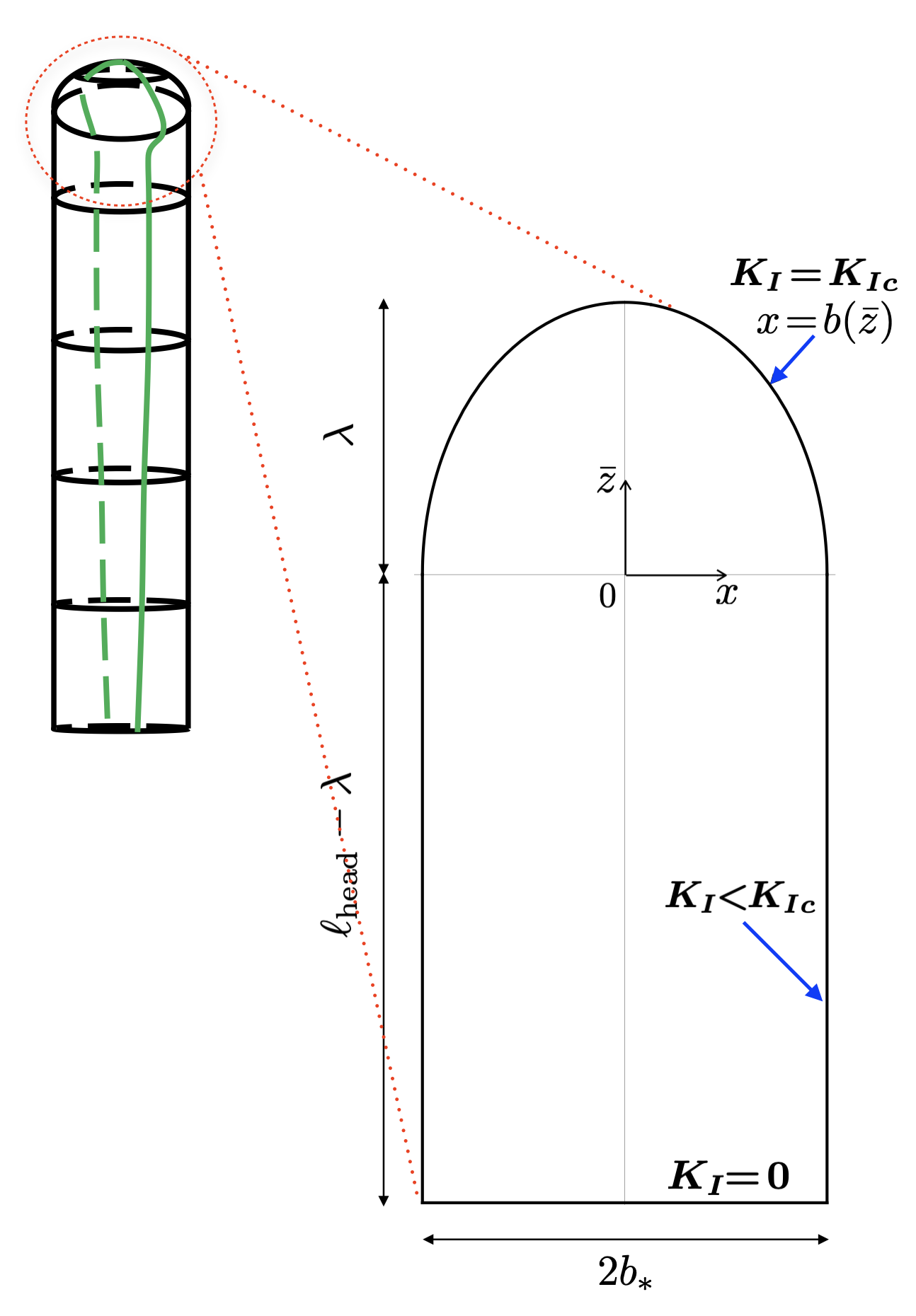}\caption{Buoyant (hydrostatic) head of a dike.\label{fig:head}}
\end{figure}

Here we look for the full solution in the fracture head (which approximate
solution we studied in the previous section), including the yet unknown
value of the stationary breadth $b_{*}$ away from the tip and the
unknown ``build-up'' $b(z)$ in the near-tip region of a priori
unknown vertical extent $\lambda$, where the fracture breadth gradually
expands towards the terminal value $b_{*}$ to accommodate the vertical
propagation (Figure \ref{fig:head}). We further make use of local
coordinate $\bar{z}=z-(\ell-\lambda)$ with the origin situated at
the boundary $\bar{z}=0$ between the laterally expanding ($\bar{z}>0$,
$b(\bar{z})<b_{*}$) and laterally stationary ($\bar{z}<0$, $b(\bar{z})=b_{*}$)
parts of the fracture edge.

Using the scales (\ref{norm}), the normalized form of the elasticity
equation (\ref{elast}) is

\begin{equation}
p(x,z)=-\frac{1}{8}\int_{\ell_{\mathrm{head}-\lambda}}^{\lambda}d\bar{z}'\int_{-b(\bar{z}')}^{b(\bar{z}')}\frac{w(x',\bar{z}')}{[(x-x')^{2}+(\bar{z}-\bar{z}')^{2}]^{3/2}}dx'\label{elast-3}
\end{equation}
while the net pressure is hydrostatic in the head
\begin{equation}
p(x,z)=\bar{z}+p_{0}\qquad(x,\bar{z})\in\text{crack footprint}\label{hydro}
\end{equation}
where $p_{0}$ is yet unknown constant. The normalized conditions
along the expanding, (\ref{wn}), and the stationary parts of the
fracture edge are, respectively, 
\begin{equation}
w_{\perp}(r\rightarrow0)=\frac{4}{\pi}\sqrt{r}\qquad(0<\bar{z}<\lambda)\label{exp}
\end{equation}
 
\begin{equation}
b(\bar{z})=b_{*}\quad\text{and}\quad w(x\rightarrow b_{*},\bar{z})<\frac{4}{\pi}\sqrt{b_{*}-x}\qquad(\ell_{\mathrm{head}}-\lambda<\bar{z}<0)\label{non-exp}
\end{equation}
where $w_{\perp}(r)$ is, as previously, the opening distribution
in the cross-section normal to the fracture edge. Finally, these are
complemented by the condition of smooth fracture edge, specifically
at the juncture $\bar{z}=0$,
\begin{equation}
db/d\bar{z}=0\quad\text{at}\quad\bar{z}=0\label{b0}
\end{equation}

We use the piecewise-constant displacement discontinuity method, as
further detailed in Appendix \ref{App:num}, to numerically solve
elasticity integral equation (\ref{elast-3}) together with constraints
(\ref{exp}-\ref{b0}) for the terminal fracture breadth and the reference
pressure value, respectively,
\begin{equation}
b_{*}\approx0.396,\qquad p_{0}\approx1.344,\label{alpha-1}
\end{equation}
the crack opening $w(x,\bar{z})$ show on Figure \ref{fig:whead},
and the fracture breadth distribution 
\[
b(\bar{z}\ge0)\approx b_{*}\sqrt{1-A(\bar{z}/\lambda)^{2}-(1-A)(\bar{z}/\lambda)^{4}}\quad\text{with}\quad A\approx0.6967
\]
in the laterally-expanding part (of the head region) of extent 
\[
\lambda\approx0.552
\]
The fracture breadth is stationary, $b(\bar{z}<0)=b_{*}$, in the
remainder of the head region of extent 
\[
\ell_{\mathrm{head}}-\lambda\approx1.504
\]
The true extent of the head region is therefore 
\begin{equation}
\ell_{\mathrm{head}}\approx2.055\label{lhead-1}
\end{equation}

We note that the net pressure at the tail-end of the head, i.e. at
$\bar{z}=-(\ell_{\mathrm{head}}-\lambda)$, has a negative value $\approx-0.1$,
which owes to the non-local 3D crack effects there. (Departure from
the PKN assumption, which would have required zero net-pressure there).

The volume of the head is 
\begin{equation}
V_{\text{head}}\approx0.408\label{Vhead-1}
\end{equation}

\subsection{PKN-Approximation of the Head}

\begin{figure}[tb]
\begin{center}
$\begin{array}[c]{r@{\hspace{.2in}}c}

\includegraphics[height=20pc]{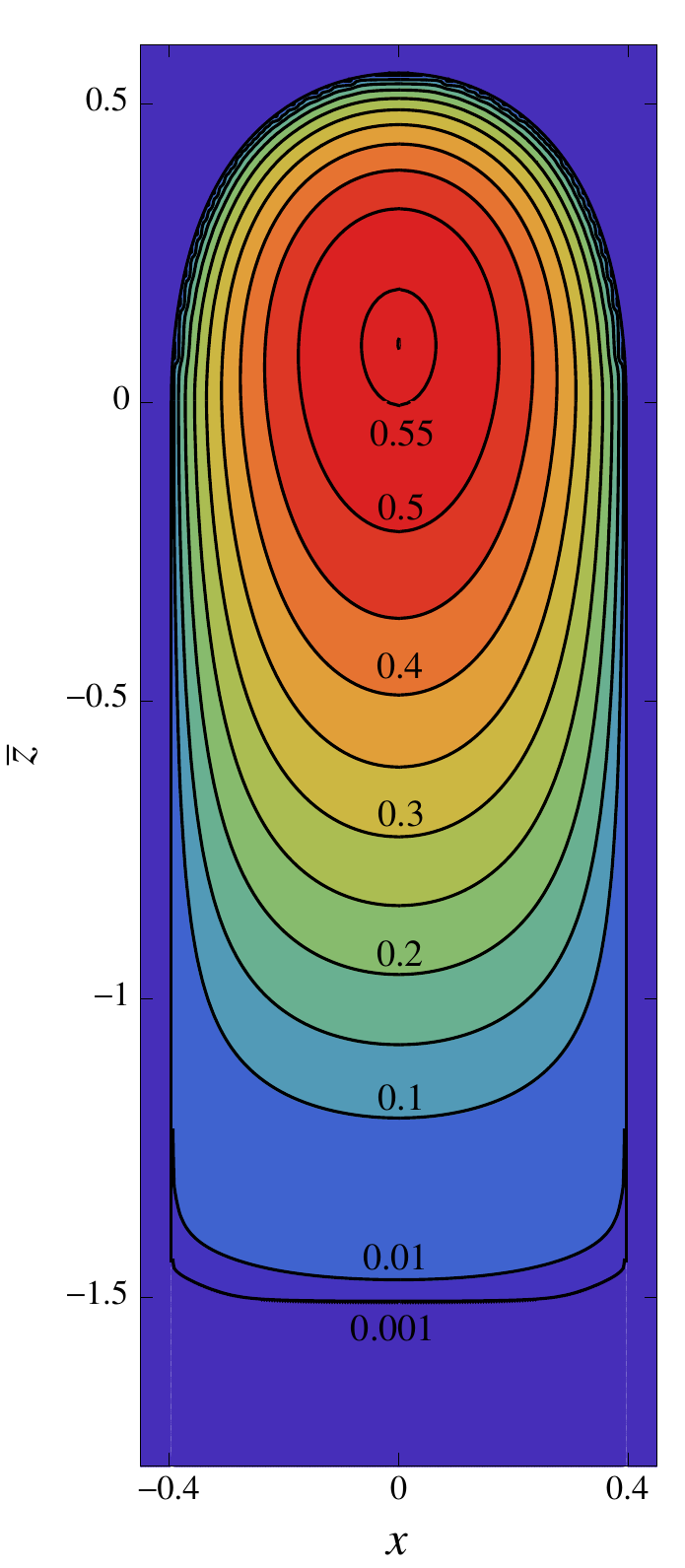} \hspace{.2in}
\includegraphics[height=20pc]{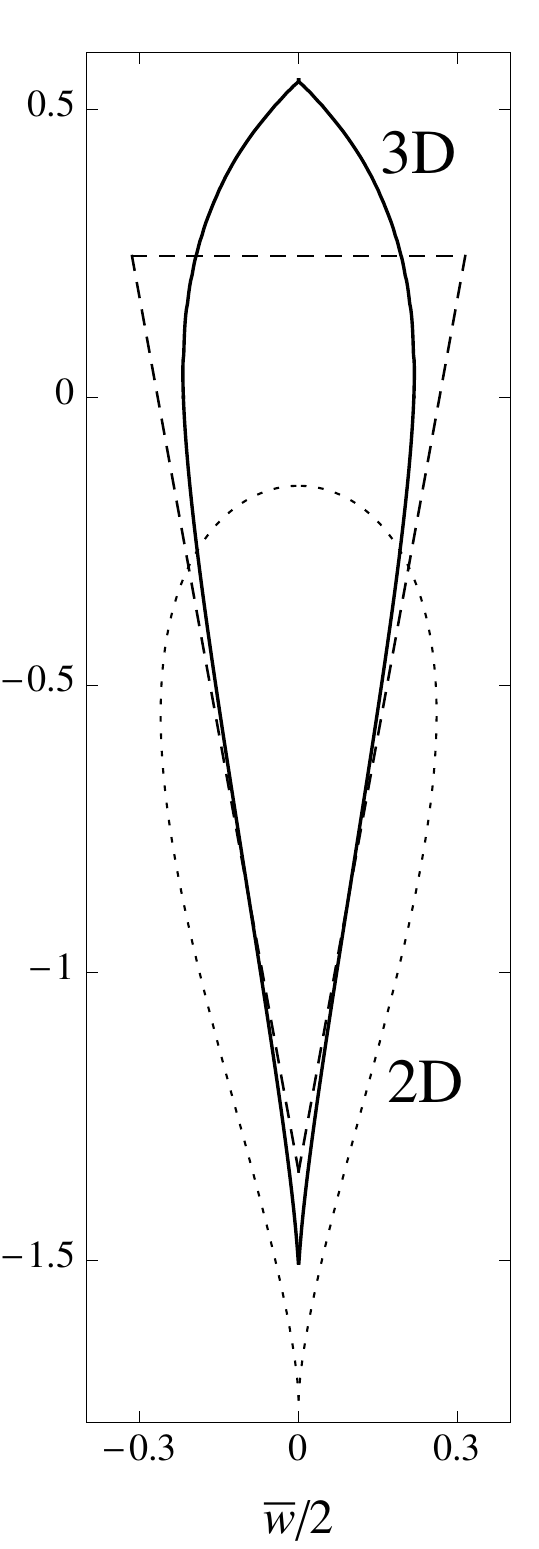} \hspace{.2in}
\includegraphics[height=20pc]{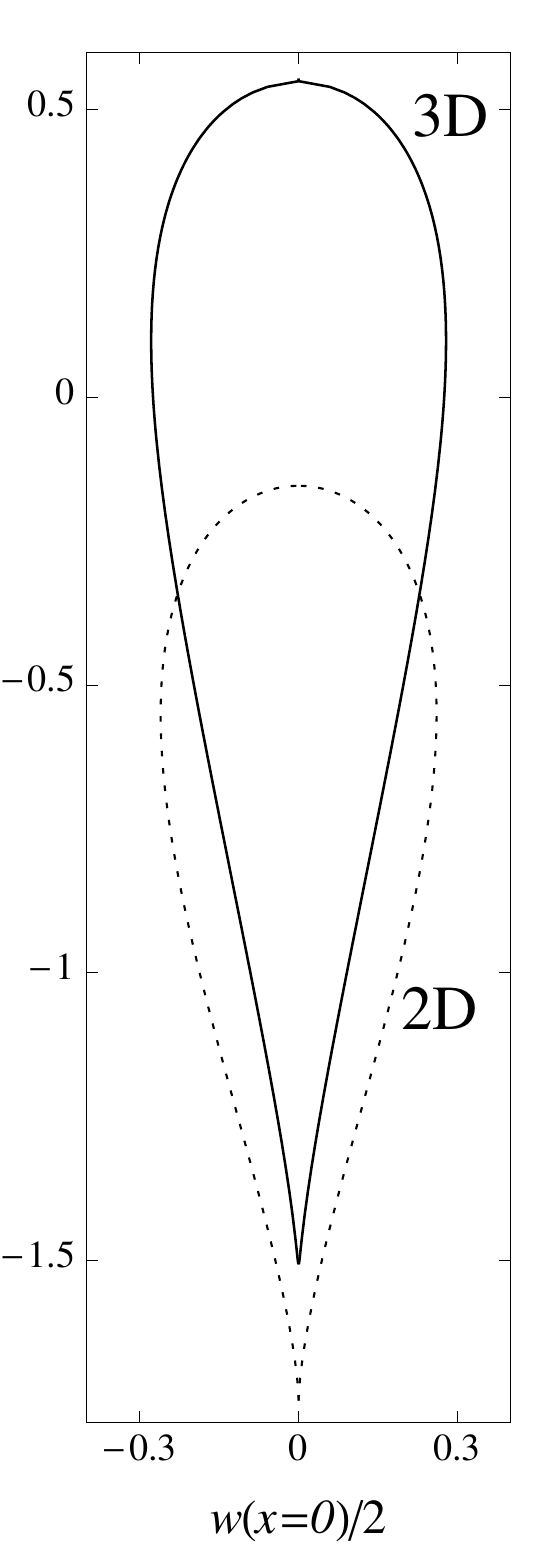} \hspace{.2in}
\includegraphics[height=20pc]{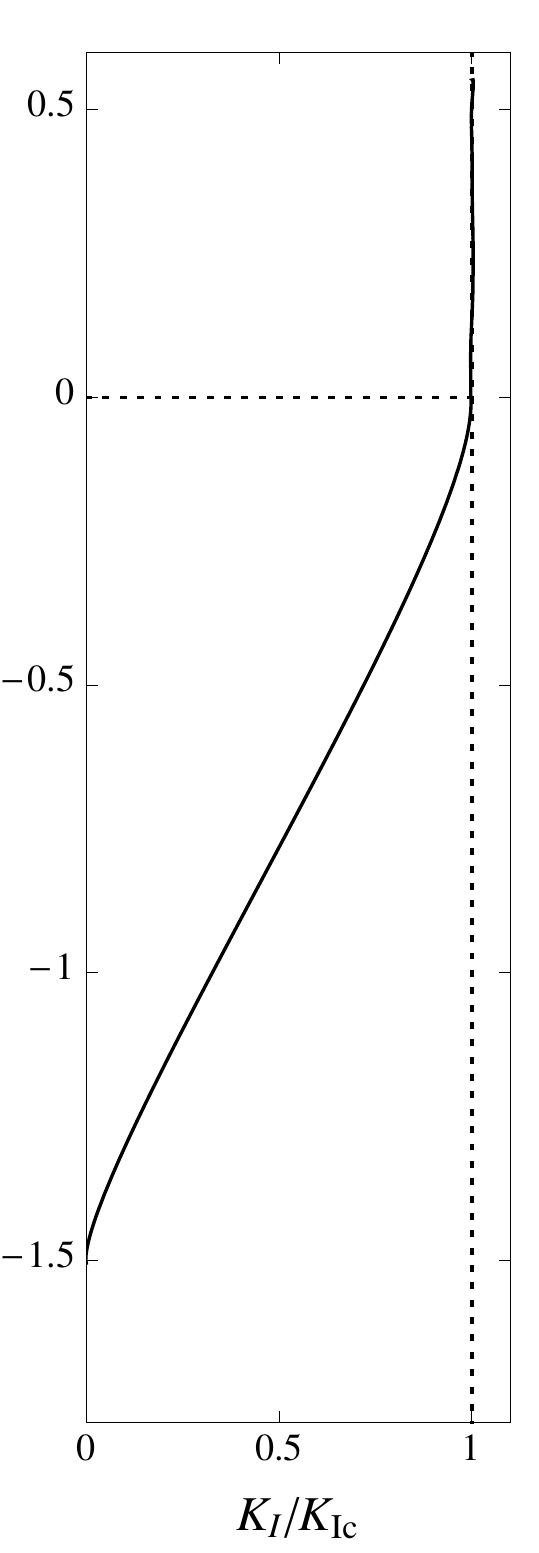}&\\[-0.65cm] 

\end{array}$ \end{center}\caption{Solution for the buoyant, hydrostatic fracture head with breadth developement.
(a) Opening $w(x,\bar{z})$, (b) Breadth-averaged opening $\overline{w}(\bar{z})$,
its PKN-approximation (dashed line), and the corresponding Weertman's
2D fracture head solution (dotted line), (c) Opening along the crack
line of symmetry $w(x=0,\bar{z})$, (d) ratio of the stress intensity
factor at the fracture edge to the toughness, $K_{I}/K_{Ic}$, (this
ratio $\approx1$ in the tip region with the expanding breadth and
$<1$ away from the tip where the breadth is stationary). The space
and the opening are scaled by the buoyancy lengthscale $L_{*}=(\bar{K}/\Delta\rho g)^{2/3}$
and $w_{*}=(\bar{K}/\bar{E})\sqrt{L_{*}}$, (\ref{norm}), respectively.
The terminal half-breadth of the crack in these units is $b_{*}\approx0.396$.
\label{fig:whead}}
\end{figure}

The PKN-approximation of the solution in the head, as shown on on
Figure \ref{fig:whead}b by dashed line, corresponds to $\overline{w}^{\mathrm{PKN}}=b_{*}p$,
where $p$ in the head is taken from the full head solution (\ref{hydro})
with (\ref{alpha-1}). The PKN approximation of the advancing front
location $\bar{z}_{\mathrm{front}}^{\mathrm{PKN}}\approx0.245$ follows
from the propagation condition $p(\bar{z}_{\mathrm{front}}^{\mathrm{PKN}})=1/\sqrt{b_{*}}$,
while the back (reseeding) front $\bar{z}_{\mathrm{back}}^{\mathrm{PKN}}\approx-1.344$
corresponds to the closure condition $p(\bar{z}_{\mathrm{back}}^{\mathrm{PKN}})=0$.
Although the validity of the PKN assumption is at best questionable
in the head region, which extent $\ell_{\mathrm{head}}$ is only about
2.5 times larger than its terminal breadth $2b_{*}$, (\ref{lhead-1}),
the PKN solution for the breadth-averaged opening starts to track
the full numerical solution some distance behind the tip (Figure \ref{fig:whead}b).
What is however even more remarkable is that the equivalent PKN volume
of the head region $V_{\mathrm{head}}^{\mathrm{PKN}}=b_{*}\approx0.396$
(see (\ref{Vhead}) and (\ref{alpha-1})), is in very good agreement
($\sim3$\% different) with the full numerical solution (\ref{Vhead-1}).
This lands support to the PKN solution for the finger crack (including
the hydrostatic head, viscous tail, and the transition between the
two) developed in Section \ref{sec:PKN}, as long as the head region
is at least partially developed, see constraints (\ref{Q_pkn}) for
the constant injection rate and (\ref{t_pkn}) for the constant volume
release cases.

\subsection{Comparison to the Weertman's 2D Head Solution}

Draw comparisons to Weertman's 2D pulse solution \citep{Weertman71}....
\[
w=\frac{2\Delta\rho g}{E'}L_{c}^{2}\left(1+\frac{Z}{L_{c}}\right)^{3/2}\left(1-\frac{Z}{L_{c}}\right)^{1/2},\qquad p=\Delta\rho g\left(Z+L_{c}/2\right)
\]
where 
\[
L_{c}=\frac{L_{pulse}}{2}=\left(\frac{K_{Ic}}{\sqrt{\pi}\Delta\rho g}\right)^{2/3}
\]
is the half-length of the pulse ($Z=\pm L_{c}$ correspond to advancing/reseeding
tips in the coordinate system used in the Weertman's solution). In
our notation, $L_{c}=L_{*}/2^{1/3}$ where $L_{*}=(\bar{K}/\Delta\rho g)^{2/3}$
is a buoyancy length, $\bar{K}=\sqrt{2/\pi}K_{Ic}$ and $\bar{E}=E'/\pi$.

Using scaling (\ref{norm}), i.e. scales $L_{*}$, $w_{*}=\bar{K}\sqrt{L_{*}}/\bar{E}$,
and $p_{*}=\bar{K}/\sqrt{L_{*}}$, we have 
\[
w=\frac{2^{1/3}}{\pi}\left(1+2^{1/3}Z\right)^{3/2}\left(1-2^{1/3}Z\right)^{1/2},\qquad p=Z+2^{-4/3}
\]
and non-dimensional pulse length is $2\times2^{-1/3}=2^{2/3}$. 

 In order compare the 2D and 3D pulse solution, we need to relate
the corresponding coordinate frames, i.e., $Z$ and $\bar{z}$, respectively.
We choose to do so by requiring that the net pressure distributions
in the two solutions are identical, i.e., $Z+2^{-4/3}=\bar{z}+p_{0}$,
which leads to $Z=\bar{z}+\left(p_{0}-2^{-4/3}\right)\approx\bar{z}+0.947$.

The normalized volume of the Weertman's 2-D pulse is equal to $0.5$,
which approximates well the equivalent 2-D-volume of the 3-D pulse,
$V_{\text{head }2D}=V_{\text{head }}/2b_{*}\approx0.515$. This result,
taken together with the approximate correspondence of the 2-D and
3-D viscous tail solutions obtained earlier, suggests that the 2-D
framework provides a good approximation (within 10\% in terms of the
dike length, width, and head vs. tail volume partition) of the 3-D
dike solution as long as the appropriate (obtained from the 3-D hydrostatic
head solution) value of the stationary breadth $b_{*}$ is used to
convert the dike's volume to its 2-D proxy.

\subsection{Comparison with Laboratory Experiments}

\begin{figure}[H]
\begin{centering}
\includegraphics[scale=0.3]{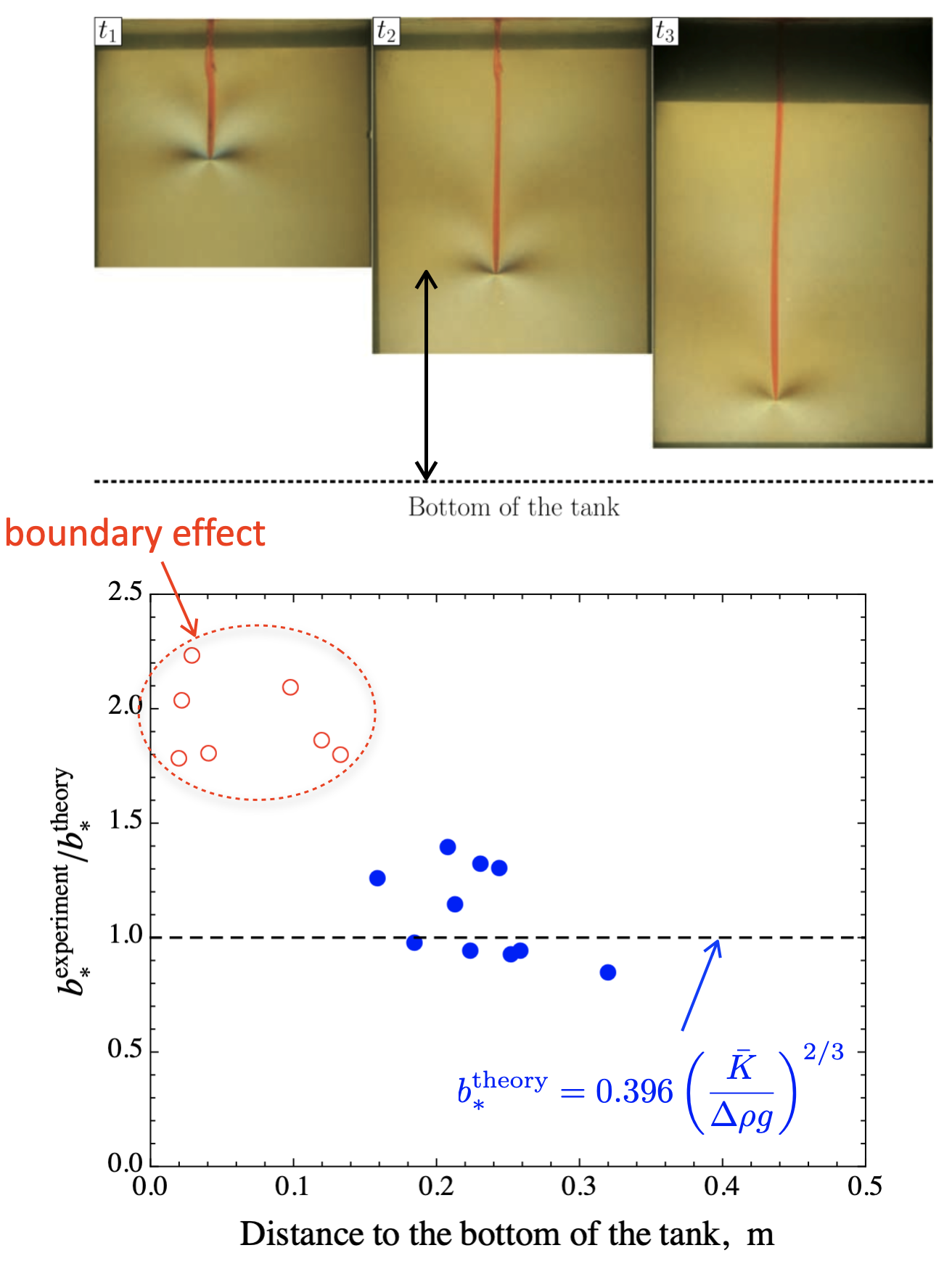}\includegraphics[scale=0.3]{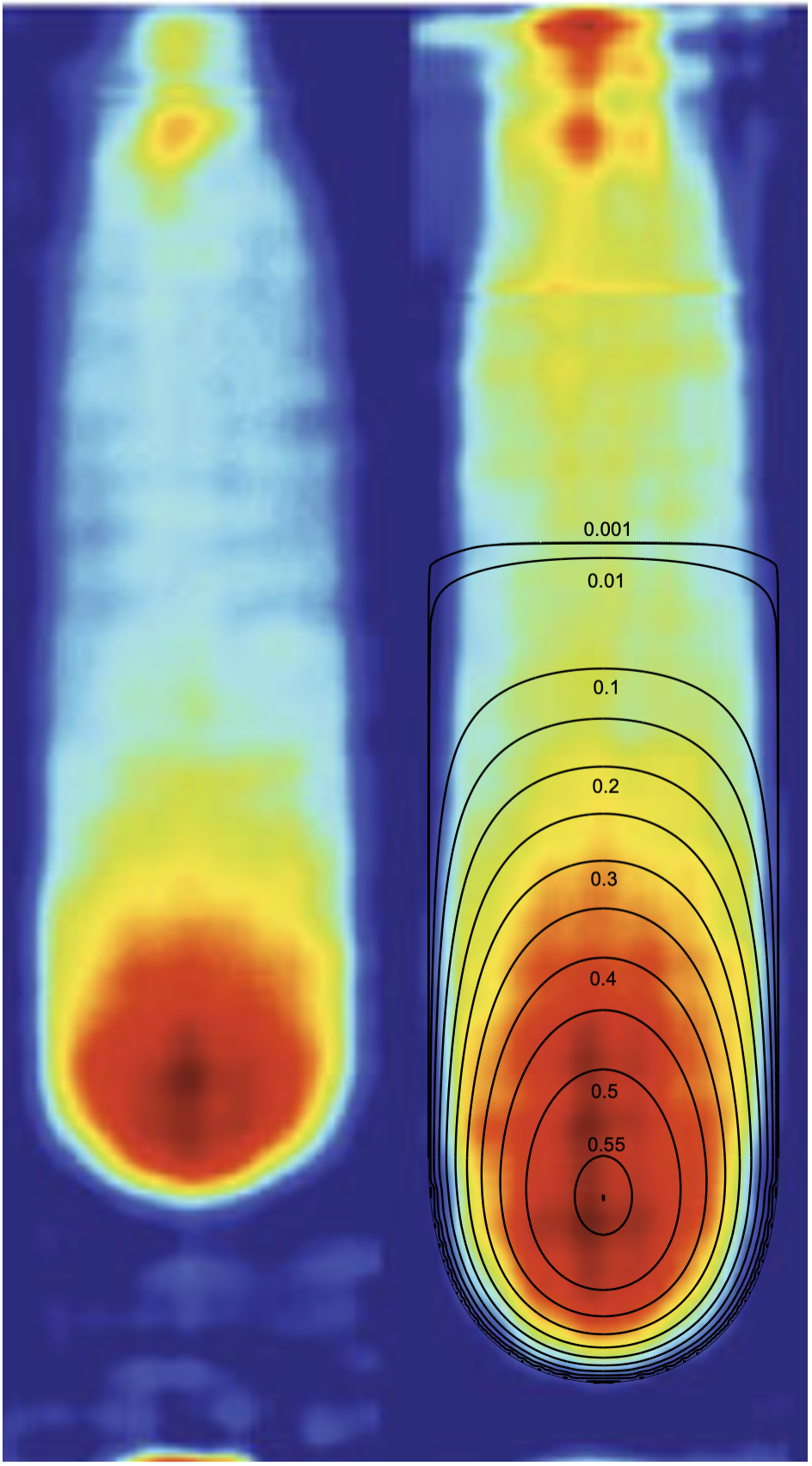}
\par\end{centering}
\caption{\textbf{(a)} Propagation of an experimental `reverse' ($\rho_{f}>\rho_{s}$)
dikes in gelatin towards the bottom boundary of the gelatin tank,
view along the dike plane (opening vs. length), adopted from Fig.
5 of \citet{TaisneTait09}. In experiments, dike eventually arrests
at some distance from the boundary. Final, arrested dike length and
breadth are reported in Table 1 of \citet{TaisneTait09}. \textbf{(b)}
Breadth of the experimental gelatin 'dikes' normalized by the theoretical
prediction $b_{*}=0.396(\bar{K}/\Delta\rho g)^{2/3}$ vs. the distance
of the arrested tip from the boundary. The agreement of the experimental
and theoretical values of the breadth deteriorates for dikes which
propagated closer to the tank boundary due to possibly apparent toughnening
in the presence of the boundary. \textbf{(c)} Not quantifie color
map of light absorption by a dike in gelatin, view normal to the dike
plane (breadth vs. length), which can serve as a proxi for the spatial
distribution of dike opening, adopted from Fig. 2 of \citet{taisne2011-dike-breadth}.
No color map scale is avaialble. Contour lines of the opening in the
theoretical solution for the 3D hydrostatic dike head (Fig \ref{fig:whead}a)
are overlayed on the experimental image for qualitative comparison.
\label{fig:TT}}
\end{figure}

\section{Application to Dikes with Fixed Volume}

We consider an example of dike driven by a fixed volume of fluid $V$
released at the source over some time $t_{inject}$. Dike dynamics
is described by the asymptotic head-tail solution discussed in the
above for $t>t_{inject}+\Delta t_{lateral}$, if the release volume
is larger than the volume accommodated in the dikes head, i.e. $V>V_{head}$
and thus 
\[
\text{sustained propagation:}\quad V_{tail}=V-V_{head}>0
\]
where head's volume is fully defined by the solids property and buoyancy
gradient
\[
V_{head}\approx0.408\frac{\bar{K}^{8/3}}{\bar{E}(\Delta\rho g)^{5/3}}
\]
Otherwise, i.e. when $V<V_{head}$, dikes is expected to arrest in
the neighborhood $\sim\ell_{head}$ of the source. 

Dike dynamics is governed by the viscosity-dominated, slowing growth
of the dike's tail
\[
\ell_{tail}\approx2.2\left(\frac{V_{tail}^{2}(\Delta\rho g)^{7/3}t}{\bar{\mu}\bar{K}^{4/3}}\right)^{1/3}
\]
This predicts slowing, but not arresting dynamics, with the ascent
rate scaling with the tail's volume given by the excess of the injected
volume over the volume of the head. In reality, eventual dike arrest
can be precipitated by changes of the buoyancy condition in the shallower
crust, and by the effective loss of dike's fluid volume to freezing
in the thinning tail.

\section{Conclusions}

A particular solution for a 3D (finite breadth) gravity driven hydraulic
fracture is developed. Solution has a similar structure to that of
2D (infinite breadth) dikes, consisting of:
\begin{itemize}
\item an extensive \textquotedblleft tail\textquotedblright , stretching
and thinning in time, with a stationary breadth, which solution is
dominated by the viscous fluid flow and characterized by negligible
elastic interactions 
\item a compact hydrostatic \textquotedblleft head\textquotedblright{} dominated
by the rock toughness and elastic interactions, a 3D analog of \textquotedblleft Weertman
pulse\textquotedblright , which determines the terminal fracture breadth 
\end{itemize}
Dyke propagation dynamics is governed by by the viscous fluid flow
in a thin tail, i.e. \textquotedblleft tail wags the dog\dots \textquotedblright{}
\citep{Stevenson82}. Yet, for 3D dykes, \textquotedblleft dog\textquoteright s
head\textquotedblright{} plays a crucial role in the dynamics, as
it creates the fracture breadth, and thus directly impacts the tail
solution and dyke dynamics by partitioning the injected fluid volume
between the head and the tail and by constraining the geometry of
the tail, the dynamics of the fluid flow therein, and thus the propagation
of the dike. 

Contrary to some previous suggestions, fixed-volume-release dykes
have slowing dynamics, but do not arrest, unless encounter tougher
or softer enough ground resulting in the head-volume expansion accommodating
entire fracture fluid volume, i.e. completely depleting dyke\textquoteright s
tail.

\bibliographystyle{plainnat}
\bibliography{bibrefs_garagash_Aug21}

\appendix

\section{PKN-like Gravity Fracture for Depth-Dependent Rock Properties\label{app:depth}}

Let us nondimensionalize the field variable with respect to the scales
\begin{equation}
x_{*}=z_{*}=L_{*}\quad t_{*}=\frac{L_{*}}{v_{*}}\quad w_{*}=\frac{\bar{K}_{*}\sqrt{L_{*}}}{\bar{E}_{*}}\quad p_{*}=\frac{\bar{K}_{*}}{\sqrt{L_{*}}}\quad v_{*}=\left(\frac{\bar{K}_{*}}{\bar{E}_{*}}\right)^{2}\frac{p_{*}}{\bar{\mu}}\quad V_{*}=w_{*}L_{*}^{2}\label{norm-1}
\end{equation}
where 
\begin{equation}
L_{*}=\left(\frac{\bar{K}_{*}}{\Delta\rho_{*}g}\right)^{2/3}\label{Lb-1}
\end{equation}
is the buoyancy lengthscale \citep{ListerKerr91} expressed in terms
of reference values $\Delta\rho_{*}$, $\bar{K}_{*}$, and $\bar{E}_{*}$
of the density mismatch $\Delta\rho=\Delta\rho_{*}\mathscr{B}(z)$,
the rock toughness $\bar{K}=\bar{K}_{*}\mathscr{K}(z)$, and the rock
modulus $\bar{E}=\bar{E}_{*}\mathscr{E}(z)$, respectively. (Functions
$\mathscr{B}$, $\mathscr{K}$, and $\mathscr{E}$ define the depth-dependence
of the rock properties).

The normalized elasticity equation (\ref{wbar}), the fluid continuity
and the Poiseuille law can be written in units of (\ref{norm-1}),
respectively, as
\begin{equation}
\overline{w}=\frac{b\,p}{\mathscr{E}}\qquad\frac{\partial b\overline{w}}{\partial t}=-\frac{\partial b\overline{q}}{\partial z}\qquad\overline{q}=-\overline{w}^{3}\left(\frac{\partial p}{\partial z}-\mathscr{B}\right)\label{PDE-2}
\end{equation}
Plugging the first and the third into the second
\begin{equation}
\frac{b^{2}}{\mathscr{E}}\frac{\partial p}{\partial t}=\frac{\partial}{\partial z}\left[\frac{b^{4}p^{3}}{\mathscr{E}^{3}}\left(\frac{\partial p}{\partial z}-\mathscr{B}\right)\right]\label{PDE_p-1}
\end{equation}
This non-linear PDE is subjected to the following global and boundary
conditions:
\begin{equation}
V=2\int_{0}^{\ell}\frac{b^{2}p}{\mathscr{E}}dz\quad\text{or}\quad2(b\bar{q})_{|z=0}=\frac{dV}{dt}\label{V-3}
\end{equation}
\begin{equation}
(\overline{q}/\bar{w})_{|z=\ell}=\frac{d\ell}{dt}\quad\text{and}\quad p_{|z=\ell}=\frac{\mathscr{K}}{\sqrt{b}}\label{tip-1-2}
\end{equation}
The normalized solution $p(z,t)$ and $\ell(t)$ of (\ref{PDE_p-1}-\ref{tip-1-2})
depends on the normalized injected fluid volume $V(t)$, normalized
buoyancy $\mathscr{B}(z)$, elasticity $\mathscr{E}(z)$, and toughness
$\mathscr{K}(z)$ parameters, and the normalized fracture breadth
$b(z)$. 

\section{Numerical Method of Lines for Solution of PKN-like Fracture\label{App:PKN}}

Non-linear PDE (\ref{PDE_p}) is to be solved with the b.c. (\ref{V})
and (\ref{tip-1}) for the evolution of the pressure $p(z,t)$ over
the crack $0\le z\le\ell(t)$ with a priori unknown length $\ell(t)$.
To avoid dealing explicitly with an unknown spatial domain, we normalize
the domain to the interval $[0,1]$ by passing to the normalized coordinate
\[
\zeta=z/\ell(t)
\]

The governing equations translate to 
\begin{equation}
\frac{\partial p}{\partial t}=\frac{\dot{\ell}}{\ell}\zeta\frac{\partial p}{\partial\zeta}-\frac{1}{b_{*}\ell}\frac{\partial\bar{q}}{\partial\zeta},\qquad\bar{q}=-b_{*}^{3}\left(\frac{1}{4\ell}\frac{\partial p^{4}}{\partial\zeta}-p^{3}\right)\label{PDE_p-2}
\end{equation}
where the partial time derivative is now understood to be taken at
a fixed $\zeta$ and $\dot{\ell}=d\ell/dt$ is the crack tip velocity.
Fluid flow boundary conditions read
\begin{equation}
V=2b_{*}^{2}\ell\int_{0}^{1}pd\zeta\quad\text{or/and}\quad\bar{q}_{|\zeta=0}=\frac{Q}{2b_{*}}\quad\text{or/and}\quad\dot{\ell}=(\overline{q}/\bar{w})_{|\zeta=1}=-b_{*}^{2}\left(\frac{1}{3\ell}\frac{\partial p^{3}}{\partial\zeta}-p^{2}\right)_{|\zeta=1}\label{V-3-1}
\end{equation}
where only two of the three conditions are independent. The fracture
propagation condition reads
\begin{equation}
p_{|\zeta=1}=\frac{1}{\sqrt{b_{*}}}\label{tip-1-2-1}
\end{equation}

Round-up of the numerical scheme: 
\begin{itemize}
\item Discretize the $\zeta\in[0,1]$ interval in $n+1$ equally-spaced
points (including the end points, $\zeta_{1}=0$ and $\zeta_{n+1}=1$).
In view of the propagation condition (\ref{tip-1-2-1}), which prescribes
the pressure at the last node, $p_{n+1}(t)=1/\sqrt{b_{*}}$, we looking
to determine the evolution in time of the $n+1$ unknowns (pressure
at $n$ grid points $p_{i}(t)$, $i=1,...n$, and the fracture length
$\ell(t)$).
\item Evaluate lubrication equation (\ref{PDE_p-2}) at the internal grid
points using the second-order space finite differences. This leads
to the $n-1$ ODEs for the pressure evolution at the internal grid
points, i.e. $\dot{p}_{i}=F_{i}(p_{1},...,p_{n+1},\ell)$, $i=2,...n-1$.
\item Evaluate two of the three relations (\ref{V-3-1}): (i) global fluid
volume balance, (ii) flow rate at the inlet, and (iii) flow velocity
at the tip (equal to the crack tip velocity). Default choices are
the rate of (i) expressed as an ODE for the pressure at the inlet,
$\dot{p}_{1}=F_{1}(p_{1},...,p_{n+1},\ell)$, and (iii), which provides
an ODE for the crack length. Different choices are used for large-time
shut-in problems, namely, the rate of (ii) with $Q=0$, is used to
evaluate the pressure rate at the inlet, and the rate of (i) is now
used to evaluate the fracture velocity. The latter change is necessitated
by the deterioration of the accuracy of the finite difference employed
in (iii) to evaluate the fluid velocity at the crack tip, as fracture
slows down and conditions in the tip region approach that in the hydrostatic
head asymptote.
\end{itemize}
The resulting system of $n+1$ ODEs is solved starting from an artificial
initial state $p_{i}(t_{ini})=1/\sqrt{b_{*}}$, $i=1,...n$, and $\ell(t_{ini})=V(t_{ini})/(2b_{*}^{3/2})$
at small, but non-zero $t=t_{ini}$, that is consistent with the propagation
condition and the global fluid balance. The impact of the choice of
$t_{ini}$ is negligible on the solution for $t\gg t_{ini}$. 

As discussed in the main text, the finger-like buoyant fracture develops
an asymptotic structure at large time consisting of the stationary
hydrostatic head, expanding viscous tail, and, in the case of the
shut-in (i.e. fixed fluid volume release), another hydrostatic region
at the crack inlet. The former and latter boundary regions are characterized
by stationary ($\ell_{\text{head}}$) and decreasing ($\ell_{inlet}\sim1/t$)
sizes, respectively, and, thus, eventually become small compared to
growing body (``viscous tail'') of the fracture. This necessitates
using a rather large number of spatial nodes in order to resolve these
two boundary layers and the overall fracture solution at large time. 

\section{Displacement Discontinuity Methodology for Hydrostatic Fracture Head
Solution\label{App:num}}

We use the piecewise-constant displacement discontinuity method to
solve elasticity equation (\ref{elast-3}) at the collocation points
located at the centers of the uniformly sized rectangular grid elements
\citep[e.g.][]{PeDe08}. Let us use a pair of indices '$kl$' to refer
to the rectangular boundary element $\{|x-x_{k}|<\Delta x/2,|\bar{z}-\bar{z}_{l}|<\Delta z/2\}$
with the center at $\{x_{k},\bar{z}_{l}\}$ and dimensions $(\Delta x,\Delta z)$,
and characterized by constant opening $w(x,\bar{z})=w_{kl}$. Then
elasticity equation (\ref{elast-3}) can be evaluated at the center
of a $ij$-elelment in the form 
\begin{equation}
p_{ij}=C_{ijkl}w_{kl}\qquad\label{discr}
\end{equation}
where the influence matrix $C$ is defined by 
\[
C_{ijkl}=[[c(x_{i}-x',\bar{z}_{j}-z')]_{x_{k}-\Delta x/2}^{x_{k}+\Delta x/2}]_{\bar{z}_{l}-\Delta z/2}^{\bar{z}_{l}+\Delta z/2},\qquad c(x,z)=\frac{\sqrt{x^{2}+z^{2}}}{8xz},
\]
and the summation convention over repeating indices is used. Since
pressure is distributed hydrostatically, (\ref{hydro}), we have for
the $ij$-element 
\begin{equation}
p_{ij}=\bar{z}_{j}+p_{0}\label{pij}
\end{equation}

We approximate the unknown fracture front $x=b(\bar{z})$ in the tip
region ($0<\bar{z}<\lambda$) by an oval shape 
\begin{equation}
b(\bar{z})\approx b_{*}\sqrt{1-A(\bar{z}/\lambda)^{2}-(1-A)(\bar{z}/\lambda)^{4}}\label{approx}
\end{equation}
where the maximum breadth $b_{*}$, the length of the near-tip region
$\lambda$, and parameter $A$ are parts of the solution. The form
(\ref{approx}) automatically satisfies the constraints at the two
ends of the tip region: $b(\lambda)=0$, $b(0)=b_{*}$, and $(db/d\bar{z})_{\bar{z}=0}=0$. 

To implement the propagation condition (\ref{exp}), an accurate evaluation
of the opening near the edge (the stress intensity factor) is required.
The numerical resolution of the constant DD method immediately near
the crack edge (the first few elements) is inadequate, i.e. opening
some distance away from the edge is to be used to determine the tip
asymptotics and the stress-intensity factor there. In practice, we
sample the numerical solution in the cross-section normal to a point
of interest on the fracture edge at distances $0.2\le r/b_{*}\le0.35$
from the edge, and fit the sampled opening values by a linear combination
of the first two asymptotic terms, $w_{\bot}(r)\approx(4/\pi)(\kappa_{0}r^{1/2}+\kappa_{1}r^{3/2})$.
The coefficient $\kappa_{0}$ is equivalent to the normalized stress
intensity factor $K_{I}/K_{Ic}$, and is required to be equal to the
unity along the expanding part of the fracture edge, (\ref{exp}).
Thus, in the numerical solution of (\ref{discr}-\ref{pij}), we search
for values of parameters $\lambda$ and $A$ of the ``oval-shape''
(\ref{approx}), the value of the terminal fracture breadth $b_{*}$,
and the reference pressure value $p_{0}$ which minimize $|\kappa_{0}-1|$
along the expanding part of the fracture edge, (\ref{approx}). When
carrying out the numerical solution, we actually rescale the space
with regard to the unknown terminal breadth $b_{*}$ in order to fix
the spatial domain of the solution. The unknown value of $b_{*}$
then assumes the meaning of the reciprocal of the unknown hydrostatic
pressure gradient in the appropriately rescaled (\ref{pij}).

\begin{figure}[t]

\includegraphics[height=18pc]{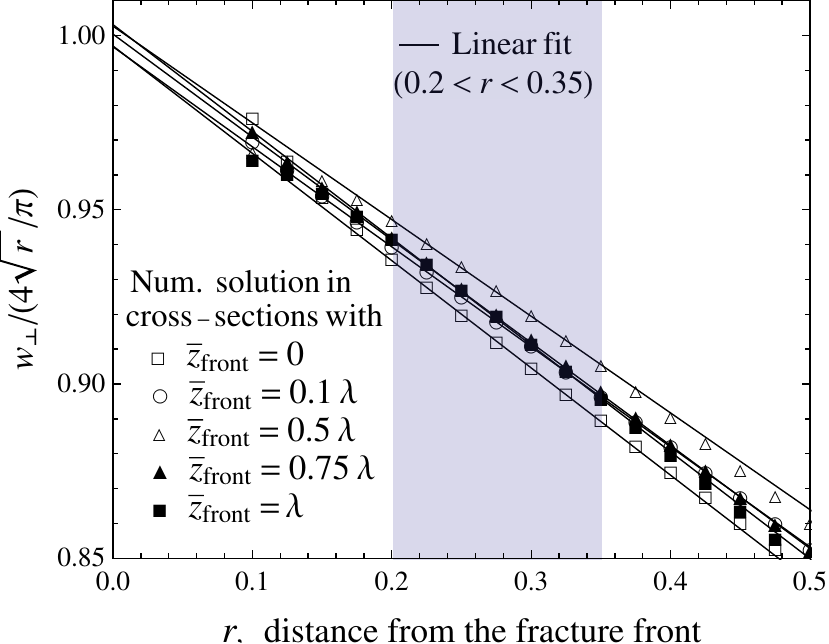} \caption{Variation of the numerical solution for the opening (normalized by
the leading tip asymptotic term), $w_{\perp}(r)/(4\sqrt{r}/\pi)$,
with distance $r$ along the normal to the fracture edge at a point
($x=b(\bar{z}_{\mathrm{front}})$, $\bar{z}=\bar{z}_{\mathrm{front}}$)
in the near tip region of the crack ($\bar{z}_{\mathrm{front}}\ge0$).
(See Figure \ref{fig:whead} for the complete solution). The linear
fit to the numerical solution sampled in $0.2<r/b_{*}<0.35$ is shown
by solid lines. The $r=0$ intercept of the fitted straight lines
provides numerical estimate of the normalized stress intensity factor
$\kappa_{0}=K_{I}/K_{Ic}$ at the corresponding point on the fracture
front. The shown case corresponds to the final solution for the buoyant
fracture head (Figure \ref{fig:whead}), and is characterized by the
minimal deviations of the $K_{I}/K_{Ic}$ from the unity in the expanding
part of the head.}

\label{fig:Kexample}
\end{figure}

The final note on the numerical implementation is related to the allocation
of the rectangular grid elements to the crack foot-print near its
curved edge, $x=b(\bar{z})$ with $\bar{z}\ge0$. The choice of these
``edge'' elements happen to have a considerable impact on the computed
values of the stress intensity factor. We set the grid element allocation
criteria by requiring the intersection area between the grid element
and the crack footprint exceeds a particular fraction of the element's
area ($\Delta x\Delta z$). To determine the optimum area fraction
we test the numerical (DD) solution for the stress-intensity factor
(determined by using the method outlined in the above) against the
analytical solution for a uniformly pressurized ($p=1$) crack with
an elliptical footprint \citep[e.g.][]{Tada00}
\[
K_{I}=\frac{\sqrt{\pi}}{\mathrm{E}(k^{2})}\left(1-k^{2}\frac{z^{2}}{\lambda^{2}}\right)^{1/4}\qquad(k^{2}=1-1/\lambda^{2})
\]
where $b=1$ and $\lambda\ge1$ are the semi-axises of the elliptical
crack and $\mathrm{E}(k^{2})=\int_{0}^{\pi/2}\sqrt{1-k^{2}\sin^{2}\phi}d\phi$
is the complete elliptic integral of the second kind \citep{AbSt64}. 

\begin{figure}[t]

\includegraphics[height=20pc]{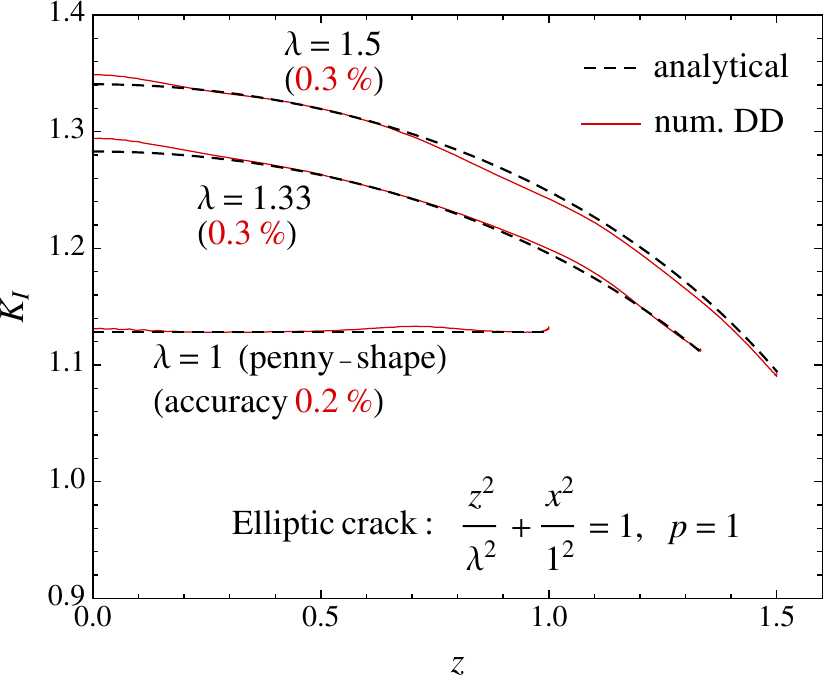} \caption{Comparison of the analytical solution for uniformly pressurized crack
with elliptical footprint to the numerical (constant DD) solution
on a rectangular grid ($\Delta x=\Delta z=0.025$), using the $77\%$
threshold for grid elements area for allocating to the crack footprint.}

\label{fig:DD-valid}
\end{figure}

We find the optimum threshold value for the grid element's area fraction
to be 0.77 (77\% of a grid element's area belongs to the crack footprint).
Figure \ref{fig:DD-valid} shows that the numerical (DD) and analytical
solutions for various values of the crack footprint aspect ratio $b/\lambda$
differ by not more than $0.3\%$. 

\end{document}